\documentclass[twocolumn]{aastex61}
\submitjournal{The Astrophysical Journal}

\shorttitle{Global current circuit structure in a resistive pulsar magnetosphere model}
\shortauthors{Yugo.E.Kato}

\usepackage{graphicx}
\usepackage{amssymb}
\usepackage{mathrsfs}
\usepackage{amsmath}
\usepackage{natbib}

\newcommand{\beq}{\begin{equation}}
\newcommand{\eeq}{\end{equation}}
\newcommand{\beqa}{\begin{eqnarray}}
\newcommand{\eeqa}{\end{eqnarray}}

\begin{document}

\title{ Global current circuit structure in a resistive pulsar magnetosphere model}
\author{Yugo.E.Kato}
\affiliation{2-4-21 Takewara, Matsuyama, Ehime, 790-0053, Japan}

\email{yugo@theo.phys.sci.hiroshima-u.ac.jp}

\begin{abstract}

Pulsar magnetospheres have strong magnetic fields and large amounts of plasma.
The structures of these magnetospheres are studied using force-free electrodynamics. 
To understand pulsar magnetospheres, discussions must include their outer region.
However, force-free electrodynamics is limited in it does not handle dissipation.
Therefore, a resistive pulsar magnetic field model is needed.
To break the ideal magnetohydrodynamic (MHD) condition $E \cdot B = 0$, Ohm's law is used.
In this work, I introduce resistivity depending upon the distance from the star and obtain a self-consistent steady state by time integration.
Poloidal current circuits form in the magnetosphere while the toroidal magnetic-field region expands beyond the light cylinder and  the Poynting flux radiation appears.
High electric resistivity causes a large space scale poloidal current circuit and the magnetosphere radiates a larger Poynting flux than the linear increase outside of the light cylinder radius.
The formed poloidal-current circuit has width, which grows with the electric conductivity. This result contributes to a more concrete dissipative pulsar magnetosphere model.

\end{abstract}

\section{Introduction}
\label{sec:intro}
Electron positron pair plasma fills a pulsar's magnetosphere, which can be described using force-free electrodynamics \cite{Goldreich1969}. A time-developed method can be used to create steady pulsar-magnetosphere solutions, as suggested by 
\citet[][]{Spitkovsky2006}, \citet[][]{Komissarov2006}, \citet[][]{2006MNRAS.368L..30M}.
Electric current flows along open magnetic-field lines whereas Poynting flux is radiated outward beyond the light cylinder. 

Force-free electrodynamics has drawbacks when the electric and magnetic fields have parallel components, causing charged particles to accelerate parallel to the magnetic field line.
Therefore, highly accelerated particles emit curvature radiation and pair creation occurs.
This process changes the pulsar's magnetospheric structure.
Deformation of the magnetosphere causes electromagnetic radiation to differ from the ideal magnetohydrodynamic (MHD) case.
Therefore, the calculation of Poynting flux using force-free electrodynamics does not use the real condition.

To address this problem, resistive electromagnetic simulation has been studied.
This method results in a clear global electromagnetic field structure, by which an amount of Poynting flux between the vacuum and force-free cases is radiated.
Electrical conductivity dependence of global current circuit structure is revealed.

In this paper, I show that such a model can produce a steady magnetosphere.
I introduce the dependence of current density on distance from the star surface.
At the outer boundary, the electrical conductivity gradually decreases, making it easy to understand a pulsar magnetosphere, including its structures and its radiation mechanism.
\subsection{Other research}

\subsubsection{Method of solving the pulsar equation}

The pulsar equation is introduced by the ideal MHD condition and the force-free approximation.
A scalar function of the pulsar equation gives the structure of the magnetosphere;
however, this equation has a singularity in the light cylinder.
The reason is that the rotational speed is limited by the speed of light.
Solving the pulsar equation in all regions has become more difficult due to the presence of the singularity.
\citet[][]{Contopoulos1999}  first solved the pulsar equation on both the inner and outer sides of the light cylinder using an iterative method to connect them. 
\citet[][]{Gruzinov2005}  examined the structure of the surrounding separatrix in the equatorial plane.
\citet[][]{Timokhin2006}  showed the solution for a different closed magnetosphere.

%%%%%%%%%%%%%%%%%%%%%%%%
\subsubsection{Time-developed method}

In quasi-analytic methods, the stability of solutions is unknown;
the time-developed method, however, can obtain stable solutions.
\citet[][]{Spitkovsky2006} found a 3D inclined rotator solution;
\citet[][]{Komissarov2006} obtained a result using a 2D axisymmetric fully relativistic MHD code, and
\citet[][]{2006MNRAS.368L..30M} did so using a force-free code.
\citet[][]{Tchekhovskoy2013} obtained a result using 3D oblique rotator fully relativistic MHD.

One weak point of this method is that electric-particle motion is neglected.
The force-free method is limited as it cannot handle pair creation.

%%%%%%%%%%%%%%%%%
\subsubsection{Particle-in-cell method}

Maxwell's equations for the electromagnetic field and charged particles constitute equations of motion, which can be solved by time integration.
The electromagnetic field is calculated at each point of the grid.
The position and velocity of the particles are calculated using the equations of motion.
A non-zero number of particles in a grid is required to represent a plasma magnetosphere.
Indeed, a large number of particles is needed to match the above conditions in all calculation regions, but computers cannot handle such numbers due to limited memory size.
Therefore, the ratio of superparticle electric charge and mass is different from that of real particles.
In particle-in-cell method, a simple pair creation model is used, and it differs for each author. 
Furthermore, the result magnetosphere strongly depends on pair creation model.

In the case without active pair creation, the pulsar magnetosphere has two charged clouds in the polar and equatorial regions.
This is known to describe a pulsar with a disc-dome configuration \citep[][]{Krause-Polstorff1985}.
This solution is not stable and experiences diocotron instability as shown by
\citet[][]{Petri2009}.

\citet[][]{2014ApJ...785L..33P} applied a sufficient pair plasma to obtain a nearly force-free aligned rotator solution. 
\citet[][]{2015MNRAS.448..606C} found the same result.
\citet[][]{Chen2014} combined pair creation and an axisymmetric magnetosphere;
their simulation did not result in pair creation. 
\citet[][]{2015ApJ...801L..19P} described a generalized-inclination rotator, whereas
\citet[][]{Cerutti2016} showed a 3D-PIC simulation.

%%%%%%%%%%%%%%%%%%%%%%%%%%%%%%
\subsubsection{Resistive Force-free Format}

The resistive force-free format adds electrical resistivity  to the force-free approximation.
In the case where electrical conductivity limit is very high, the result is expected to approach the force-free approximation.
The current density formula is derived from Ohm's law.
The time development of Maxwell equations and the current-density equation shows a steady solution.

A current density model was first applied to a pulsar magnetosphere by \citet[][]{Lyutikov2003b}.
This model is derived from Ohm's law, but has one free parameter: the velocity along the magnetic field line. 
In order to simplify this equation, \citet[][]{Lyutikov2003b} set the velocity along the magnetic field to zero.

The characteristic of this current density model gives a space-like electromagnetic field.
Some works present different results from the ideal MHD case.
Different current densities model different magnetospheric solutions.
The structure of the current sheet and the emitted Poynting flux are different.Poynting
\citet[][]{Gruzinov2007},\citet[][]{Gruzinov2008} noticed the sign of the current density four-vector.
The space-like current-density region differs from the case of ideal MHD.

\citet[][]{Li2011} shows the Poynting flux dependence of $(\sigma/ \omega)^2$. 
The poynting flux of the magnetosphere has been obtained as an intermediate value in the vacuum and force-free cases.

So far, the shape and width of the electric current have not been examined.
Therefore, I consider these properties under change to the global region of current flow.

%%%%%%%%%%%%%%%%%%%%%%
\section{Current density model}

\subsection{Force-free electromagnetic fields}

  In the case of the force-free approximation, the current density is 
uniquely determined by electromagnetic fields under the ideal MHD condition. 
The explicit form will be derived below. 
The force-free condition is written as
\begin{eqnarray}
\rho c\mathbf{E}  + \frac{\mathbf{j} \times \mathbf{B}}{c} = 0.
\label{eq:eq1}
\end{eqnarray}

By the cross product to eq. (\ref{eq:eq1}) with $\mathbf{B}$ and using a vector calculus identity, 
we have
\begin{eqnarray}
\mathbf{j} = \rho c \frac{\mathbf{E}\times \mathbf{B} }{ B^{2} } 
+ \frac{(\mathbf{j} \cdot \mathbf{B}) \mathbf{B}}{B^{2}},
\label{eq:eq2}
\end{eqnarray} 
where $ B^{2}\ne 0$ has been assumed.
We also assume that the ideal MHD condition $\mathbf{E} \cdot \mathbf{B} = 0$ 
always holds, such that the time derivative is also zero.
\begin{eqnarray}
\frac{\partial }{\partial t} (\mathbf{E} \cdot \mathbf{B} ) =
\frac{\partial \mathbf{E} }{\partial t} \cdot \mathbf{B} 
+ \mathbf{E} \cdot \frac{\partial \mathbf{B} }{\partial t } =0.
\label{eq:ebdt}
\end{eqnarray}
The time-derivative terms of the electromagnetic fields are 
substituted by the Maxwell equations, and eq. (\ref{eq:ebdt}) becomes 
\begin{eqnarray}
(\mathbf{\nabla} \times \mathbf{B} ) \cdot \mathbf{B}
 - \frac{4\pi}{c} \mathbf{j} \cdot \mathbf{B}  - 
( \mathbf{\nabla} \times \mathbf{E})  \cdot \mathbf{E} =0.
\end{eqnarray}
By eliminating the term $\mathbf{j} \cdot \mathbf{B}$ from eq (\ref{eq:eq2}), the current density is given by
\begin{eqnarray}
\frac{4 \pi}{c} \mathbf{j} =  
\frac{(\mathbf{\nabla} \cdot \mathbf{E} ) \mathbf{E} \times \mathbf{B}}{B^{2}} 
+\frac{ \mathbf{B}  \cdot (\mathbf{\nabla} \times \mathbf{B}) -\mathbf{E}  \cdot (\mathbf{\nabla} \times \mathbf{E} ) }{B^{2}}\mathbf{B},
\label{ffden}
\end{eqnarray}
where the charge density $\rho$ is replaced by Gauss's law,
$\mathbf{\nabla}  \cdot \mathbf{E}= 4 \pi \rho$.

The first term in eq.(\ref{ffden}) is the $\mathbf{E} \times \mathbf{B} $ drift.
If $|E|<|B|$, then there is a frame in which the electric field vanishes.
The velocity of this frame measured in the lab frame is 
$\mathbf{E} \times \mathbf{B}/B^2 $.
Unfortunately, the condition $|E|<|B|$ is not guaranteed in the dynamics.
Physically, if $|E|>|B|$ happens elsewhere, 
strong currents should flow to reduce the electric field. 
In the numerical calculation, this effect should be included by hand 
to construct proper force-free fields.
The second term in eq.(\ref{ffden}) describes the current flow along 
the magnetic field. 
In this way, the current density is determined uniquely for 
given $\mathbf{E}, \mathbf{B}$ in the force-free approximation.
The electromagnetic fields $\mathbf{E}$ and $\mathbf{B}$ should be solved 
with the source (\ref{ffden}) involving the spatial derivatives 
of $\mathbf{E}$ and $\mathbf{B}$.
Thus, causal structure, i.e., information propagation with light 
velocity $c$, is broken in this approximation. 

\subsection{Resistive scheme}

We here derive another current model described by electromagnetic 
fields, namely the
resistive model proposed by \citet[][]{Li2011} .
Our basic assumption is Ohm's law, which in the fluid rest frame is
\begin{eqnarray}
\mathbf{j}_{\mathrm{fluid}} \equiv \sigma \mathbf{E}_{\mathrm{fluid}},
\end{eqnarray}
where $\sigma$ is the electric conductivity.

The relation between the current density and the electromagnetic fields in 
the laboratory frame is given by the Lorentz transformation.
We consider a frame where the electric and magnetic fields are parallel.
The fluid rest frame is given by the Lorentz boost
$\mathbf{\beta}_{1}$ along the parallel.
The laboratory frame is given by another boost, $\beta_{\parallel}$,  
in the $\mathbf{E} \times \mathbf{B}$ direction.
Thus, the laboratory frame can be connected to the fluid rest frame 
through two boosts with $\beta_{1}$ and $\beta_{\parallel}$.
After some algebra, we have 
\begin{eqnarray}
\mathbf{j} = \frac{\rho_{e} c \mathbf{E} \times \mathbf{B} }{B^{2} + E_{0}^{2}} + \frac{\left( -\beta_{\parallel} \rho_{e} c + \sqrt{\frac{B^{2}+E_{0}^{2}}{B_{0}^{2}+E_{0}^{2} } (1-\beta_{\parallel}^{2}) } \sigma E_{0} \right) \left( B_{0}\mathbf{B} + E_{0}\mathbf{E} \right) }{B^{2} + E_{0}^{2}},
\label{denrf}
\end{eqnarray}
where $E_{0}$ and $B_{0}$ are given by two Lorentz invariants
defined by ($E_{0}>0$)
\begin{eqnarray}
E_{0}^{2} - B_{0}^{2} &=& \mathbf{E}^{2} -\mathbf{B}^{2},  \\
E_{0} B_{0} &=& \mathbf{E} \cdot \mathbf{B}.
\end{eqnarray}
They are explicitly solved as

\begin{eqnarray}
B_{0}^{2} &=& \frac{1}{2} \left(  \mathbf{B}^{2} -\mathbf{E}^{2} 
+ \sqrt{(\mathbf{B}^{2} - \mathbf{E}^{2})^{2} +4 (\mathbf{E} \cdot \mathbf{B})}  \right), \\
E_{0}  &=& \sqrt{B_{0}^{2} -\mathbf{B}^{2} +\mathbf{E}^{2}  },\\
B_{0}  &=& \mathrm{sign} (\mathbf{E} \cdot \mathbf{B} ) \sqrt{B_{0}^{2}}.
\end{eqnarray}
The magnitude of $\beta_{\parallel}$ is given by
\begin{eqnarray}
\beta_{\parallel} = \sqrt{\frac{B^{2} -B_{0}^{2} }{B^{2} + E_{0}^{2} }}.
\end{eqnarray}
There remains one parameter, $\beta_{\parallel}$, which
describes fluid speed in the direction of the magnetic field.

The current density from eq. (\ref{denrf})
can also be expressed by $\mathbf{E} $ and  $\mathbf{B} $ only, as in the force-free case.
By introducing $E_{0}$, the velocity in the first term
is always less than $c$. The velocity is furthermore reduced to 
the standard drift-velocity in the case where
$\mathbf{E} \cdot \mathbf{B} =0$.

The second important point is the fact that 
eq. (\ref{denrf}) is expressed by $\mathbf{E}$ and $\mathbf{B}$
without spatial derivatives
other than the charge density.
This is in contrast to the force-free case.

\subsection{Speed along the magnetic field}
We here discuss the parameter $\beta_{\parallel}$,
which describes the fluid velocity in the direction of the magnetic-field 
line. This quantity should be determined only by physical argument.
\citet[][]{Gruzinov2011} proposed that the relativistic four-current vector $( \rho_{e}, \mathbf{j})$ should be space-like.
This condition determines $\beta_{\parallel}$ as
\begin{eqnarray}
\beta_{\parallel} = 
\frac{- \rho_{e}}{ \sqrt{\gamma^{2}_{x} \sigma^{2} E_{0}^{2} + \rho_{e}^{2} c^{2}}}.
\end{eqnarray}
With this choice, eq. (\ref{denrf}) becomes
\begin{eqnarray}
\mathbf{j} = \frac{\rho c \mathbf{E} \times \mathbf{B} }{B^{2} + E_{0}^{2} } 
+ \frac{\sqrt{\gamma_{x}^{2} \sigma^{2} E_{0}^{2} + \rho^{2}_{e}c^{2}}
 (B_{0} \mathbf{B} +E_{0} \mathbf{E})}{B^{2} + E_{0}^{2} },
\end{eqnarray}
where $\gamma_{x}$ is given by
\begin{eqnarray}
\gamma_{x}^{2} = \frac{B^{2} + E_{0}^{2} }{ B_{0}^{2} + E_{0}^{2} } .
\end{eqnarray}

Space-like four current may not 
be adequate for the pulsar magnetosphere.
Within closed magnetic field lines,
charge-separated plasma co-rotates with the central star.
However, this condition may not hold in the open-field-line region.
 
Li et al.(2012) assumed $\beta_{\parallel}=0$;
i.e., the velocity parallel to the magnetic field is slowest. 
This assumption may not be justified, but
eq. (\ref{denrf}) has the simple form
\begin{eqnarray}
\mathbf{j} = \frac{\rho_{e} c \mathbf{E} \times \mathbf{B} 
+ \sqrt{\frac{B^{2}+E_{0}^{2}}{B_{0}^{2}+E_{0}^{2}}  } 
\sigma E_{0} \left( B_{0} \mathbf{B} +E_{0} \mathbf{E}  \right) }{B^{2}+E_{0}^{2}}.
\label{jmodelbeta0}
\end{eqnarray}
In our numerical calculation, we adopt this type of current model.

\subsection{Correspondence between the force-free and vacuum solutions}
When electrical conductivity $\sigma=0$ and electric charge $\rho_{e}=0$ in the vacuum;
the force-free approximation corresponds to $\sigma = \infty$. 
The toroidal field current sheet is formed on the equatorial plane in the force-free approximation, but not in the vacuum case as no current flows in the vacuum.

The amount of current flowing through the magnetosphere will vary depending on the electrical conductivity $\sigma$.
Poynting flux does not diverge in the limit of infinite electrical conductivity because there are no more vacuum gaps.

\subsection{Difference in the solution due to the current density model}
In the case of force-free electrodynamics, the form of the electric current is unique.
In the resistive force-free scheme, the expression of the current density is not determined uniquely, because the speed in the magnetic field line direction is a free parameter.
Li et al. (2012) set $\beta_{\parallel}=0$.
The region where magnetic field lines are closed is formed properly in the case of the uniform electrical conductivity.
Poloidal current flows in the region where the magnetic field lines are open.
On the other hand, different current densities in the model proposed by \citet[][]{Gruzinov2007}.
Guruzinov proposed a current density model to satisfy the space-like current density conditions.
But, it remains unclear whether this current density model is valid or not for pulsar magnetospheres.

\section{Numerical method and problem setup}%ch 4

%\subsection{Assumptions and Equations}%4.1
%(2.1)%%%%%%%%%%%%%%%%%%%%%%%%%%%%%%%%%%%%%%%%%
\subsection{Electromagnetic fields}%4.1.1
%%%%%%%%%%%%%%%%%%%%%%%%%%%%%%%%%%%%%%%%%%%%%%%
  Maxwell's equations are solved with the charge 
density $\rho _{e}$ and the current density $\mathbf{j}$:
\begin{equation}
\frac{1}{c}\frac{\partial \mathbf{B}}{\partial t}=
  -\mathbf{\nabla}\times \mathbf{E},
\label{eqn.Farad}
\end{equation}
\begin{equation}
\frac{1}{c}\frac{\partial \mathbf{E}}{\partial t}=
  \mathbf{\nabla}\times \mathbf{B}-\frac{4\pi \mathbf{j}}{c},
\label{eqn.Amp}
\end{equation}
\begin{equation}
\mathbf{\nabla}\cdot \mathbf{B}=0, 
\label{eqn.Gauss}
\end{equation}
\begin{equation}
\mathbf{\nabla}\cdot \mathbf{E}=4\pi \rho _{e}.
\label{eqn.Coulb}
\end{equation}
%%%%%%
In order to solve these equations, we use a scalar 
potential $\Phi $ and a vector potential $\mathbf{A}$ satisfying 
the Coulomb gauge, $\mathbf{\nabla}\cdot \mathbf{A}=0$. 
For axially symmetric fields, the following form given by
two functions, $F(t,r,\theta )$ and $G(t,r,\theta )$, is automatically
satisfied with the gauge condition: 
\begin{equation}
\mathbf{A}=\frac{1}{r\sin \theta }\mathbf{\nabla}F\times \mathbf{e}_{\phi }
+\left( \frac{G}{r\sin \theta }\right) \mathbf{e}_{\phi },
\end{equation}
where $\mathbf{e}_{\phi }$ is a unit vector in the azimuthal direction. 
The magnetic field $\mathbf{B}$ is given by $ \mathbf{\nabla}\times \mathbf{A}:$
%\bigskip
\begin{equation}
\mathbf{B} =\frac{1}{r\sin \theta }\mathbf{\nabla}G\times \mathbf{e}_{\phi }+\left( 
\frac{S}{r\sin \theta }\right) \mathbf{e}_{\phi },
  \label{eqn.B}
\end{equation}
where the function $S$ in $B_{\phi }$ is given by
\begin{equation}
  S=-{\mathcal{D}}F,
 \label{eqn.F}
\end{equation}
and the differential operator $\mathcal{D}$ is defined by 
\begin{equation}
{\mathcal{D}}\equiv  \frac{\partial ^{2}}{\partial r^{2}}
+\frac{\sin\theta }{r^{2}}\frac{\partial }{\partial \theta }
\left( \frac{1}{\sin \theta }\frac{\partial }{\partial \theta }\right).
\end{equation}

The electric field $\mathbf{E}$ is expressed by the
time derivative of $\mathbf{A}$ and the gradient of 
$ \Phi $:
\begin{eqnarray}
\mathbf{E} 
&= &
-\mathbf{\nabla}\Phi -
\frac{\partial }{c\partial t} \mathbf{A}
\\
&=&
-\mathbf{\nabla}\Phi -\frac{1}{r\sin \theta }\mathbf{\nabla}
\left( \frac{\partial F}{c\partial t}\right)\times \mathbf{e}_{\phi }
-\left( \frac{1}{r\sin \theta }
\frac{\partial G}{c\partial t}\right) \mathbf{e}_{\phi }.  
\label{eqn.E}
\end{eqnarray}
%

%%%
The electric field (eq.(\ref{eqn.E})) is given 
by the time derivative of $F$ such that 
it is convenient to solve the time derivative of eq.(\ref{eqn.F}),
i.e., 
\begin{equation}
 {\mathcal{D}}
\left( \frac{\partial F}{\partial t} \right) 
=-\frac{\partial S}{\partial t} .
  \label{eqn.FTM}
\end{equation}%
Substituting these forms (eqs.(\ref{eqn.B}) and (\ref{eqn.E}))
into eq.(\ref{eqn.Amp}), we have two wave
equations for $G$ and $S$.
They are the $\phi$ component and the rotation of the poloidal 
component of eq.(\ref{eqn.Amp}):
\begin{equation}
 \left( \frac{1}{c^2}\frac{\partial ^2}{\partial t^2 } - {\mathcal{D}}
  \right) G = \frac{4 \pi }{c} j_\phi r \sin \theta,
  \label{eqn.G} 
\end{equation}
\begin{equation}
 \left( \frac{1}{c^2}\frac{\partial ^2}{\partial t^2 } - {\mathcal{D}}
\right ) S = 
\frac{4 \pi}{c} \left( \frac{\partial ( r j_\theta) }{\partial r }
 - \frac{\partial j_r }{\partial \theta } \right) \sin \theta .
  \label{eqn.S}
\end{equation}
From eq.(\ref{eqn.Coulb}), we have
the Poisson equation for $\Phi$: 
\begin{equation}
 \left( \frac{1}{r^2} \frac{ \partial }{\partial r } \left( r^2 \frac{
\partial }{\partial r }\right) +\frac{1}{r^2 \sin \theta } \frac{ \partial }{%
\partial \theta } \left( \sin \theta \frac{ \partial }{\partial \theta }
\right) \right) \Phi = - 4 \pi \rho_e. 
 \label{eqn.P}
\end{equation}
One advantage to the potential formalism is that the 
constraints, eqs.(\ref{eqn.Farad}) and (\ref{eqn.Gauss}),
are automatically satisfied.

\subsection{ Region }
%%%%%%%%%%%%%%%%%%%%%%%%%%%%%%%%%%%%%%%%%%%%%%%

In our numerical calculation, we use
the spherical coordinate $(r,\theta)$
with range $ r_{0} \le r \le r_\mathrm{out}$,
and  $ 0 \le \theta \le \pi/2 $.
Typically, we set $ r_\mathrm{out}/r_{0} = 60$
The light cylinder, $R_{L}$, is located at 
$R_{L}/r_{0}=1/\Omega =5 $.
We use the finite difference method to solve
the partial differential equations.
The typical numbers of cells on the grid are $60$ and $96$ in 
the $r$ and $\theta$ directions, respectively.
The grid-cell spacing in the radial direction is taken 
as $ \Delta r \propto r^{1/2}$
to obtain fine resolution near the inner region, 
whereas the spacing in the angular direction
is constant.

\begin{figure}
\begin{center}
\includegraphics[angle=0,trim=1cm 0cm 0.5cm 0.5cm,width=0.95\columnwidth]{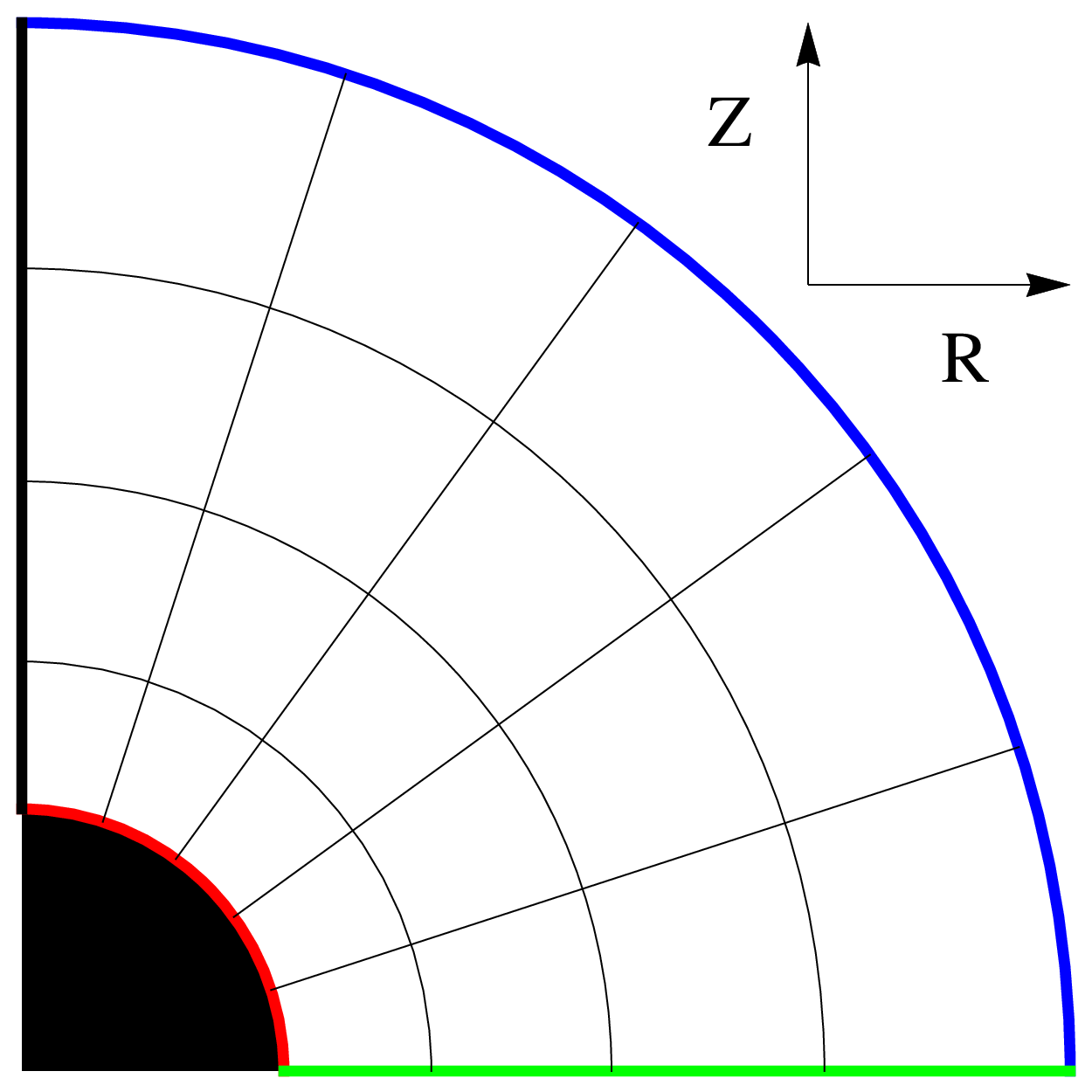}
\end{center}
\label{pic:Proggrid.pdf}
\caption{Figure of the two dimensional computational domain and four boundaries. The thin black line represents the grid. The red line is star surface, thick black line is the rotation and magnetic axis, the green line is the  equatorial plane, and the blue line represents the outer boundary.}
\end{figure}

\begin{figure}
\begin{center}
\includegraphics[angle=0,trim=1cm 0cm 0.5cm 0.5cm,width=0.95\columnwidth]{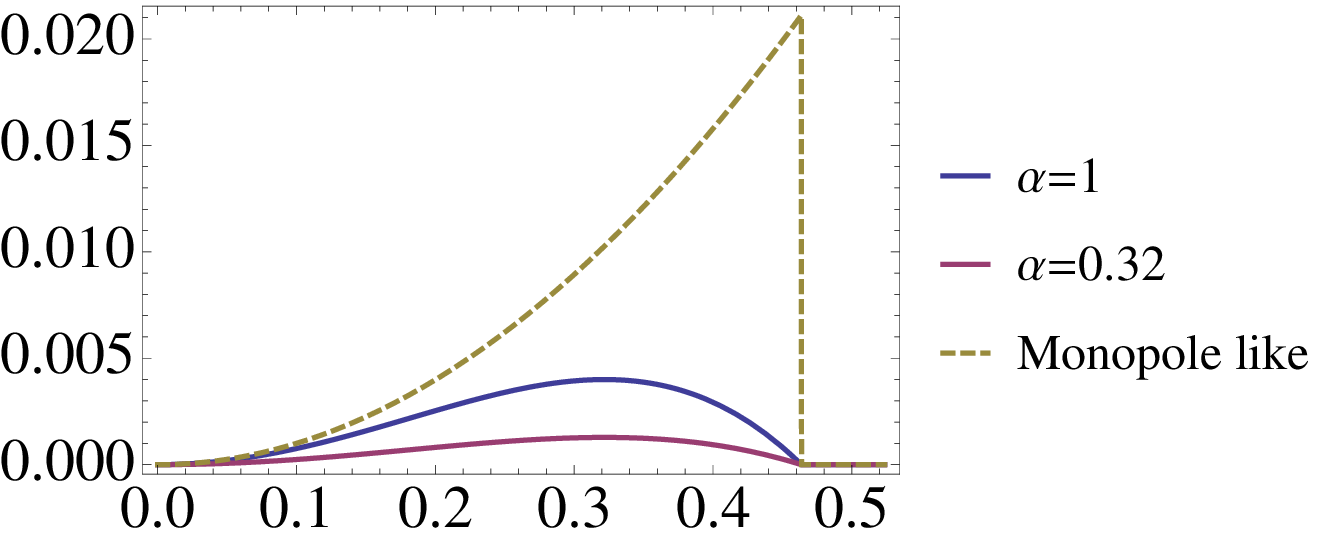}
\end{center}
\caption{The structure of the surface toroidal flux function $S_{0}(r=r_{0},\theta)$}
\label{pic:Ssuf2.pdf}
\end{figure}

\subsection{Boundary conditions}
%%%%%%%%%%%%%%%%%%%%%%%%%%%%%%%%%%%%%%%%%%%%%%%
%
In this section, we discuss the boundary conditions.
First, we consider the condition at the inner boundary $ r_{0} $.
The poloidal magnetic field there is a dipole field given by
\begin{equation}
\mathbf{B}_{d} = [B_{r}, B_{\theta }, B_{\phi } ] 
 = \left [ \frac{2 \mu \cos \theta }{r^3}, 
           \frac{ \mu \sin \theta }{r^3} , 0 \right ],      
\label{eq:kyoukai_B}
\end{equation}
where $\mu$ is the dipole moment and 
the field strength on the pole $r_{0} $ is 
$B_{0}=2\mu /r_{0}^3$.
The magnetic flux function $G$ 
for a purely dipole field
is given by
\begin{equation}
 G_{d} = \frac{\mu \sin ^2 \theta }{r} .
  \label{eqn.puredip}
\end{equation}
We always fix the function $G$ at $r=r_{0}$ as
\begin{equation}
 G(r_{0}, \theta )  = \frac{\mu \sin ^2 \theta }{r_{0}} .
  \label{eqn.puredip0}
\end{equation}
Thus, the continuity of the radial component $B_{r}$ 
is guaranteed. 
%$B_{r}, B_{\theta } $ 

%2%
Inside the surface $r \le r_{0}$, the
ideal MHD condition holds such that
the electric field is given by 
the rotational velocity  
$\mathbf{v} =(r \sin \theta )\Omega  \mathbf{e}_{\phi}   $
and the magnetic-dipole field
\begin{equation}
\mathbf{E} = -\mathbf{v} \times  \mathbf{B} _{d} 
 = \left [  \frac{ \mu \Omega \sin^{2} \theta }{r^2_{0}},
           -\frac{2 \mu \Omega\cos \theta \sin \theta}{r^2_{0}}, 0 \right ].     
%  \label{eqn.interE}
\end{equation}
The $\theta$ component is
given by $E_{\theta } = -\Omega \partial G_{d}/\partial \theta $. 
From the continuity of $E_{\theta }$,
the electric potential $\Phi$ at the surface can be chosen 
as
\begin{equation}
\Phi(r_{0}, \theta )  = 
\frac{\mu \Omega \sin ^2 \theta }{r_{0}} ,
% = \Omega G(r_{0}, \theta )  = 
% 
\label{boundphi}
\end{equation}%
and 
\begin{equation}
\frac{ \partial F}{ \partial t} =0.
\end{equation}%  

Next, we consider 
the toroidal magnetic field, $B_{\phi} = S/(r \sin\theta)$.
For the force-free case,
the current function $S$ is given by
a function of $G$.
In particular, in the split-monopole solution,
the toroidal magnetic flux is given by
\begin{equation}
S_{m} = -\Omega G_{m} \left(
2 - \frac{G_{m}}{B_{0} r_{0} ^2} \right),
%  \label{eqn.vlpert}
\end{equation}%
and $ G_{m} =B_{0} r_{0} ^2 (1-\cos \theta) $,
where $ B_{0}$ is a  constant,
$ B_{r} =B_{0} (r_{0}/r)^2$.
The toroidal magnetic flux can also be expressed as a function of
$\theta$ by eliminating $G_{m}$:
\begin{equation}
S_{m} = - \Omega B_{0} r_{0} ^2 \sin^{2} \theta .
 \label{eqn.Smono}
\end{equation}%
The current function 
should vanish for the rotating dipole in the region  
$ \theta > \theta _{p} $ at the surface,
where $\theta _{p} $ is the polar-cap angle
$\sin \theta _{p} = (\Omega r_{0})^{1/2}$.
We expect that the magnetic field near the pole,
even for a rotating dipole,
will be similar to that in the split-monopole case.
At the same time, the functional form (\ref{eqn.Smono}) 
should be truncated for $ \theta > \theta _{p} $.
Our choice of $S$ at $r_{0}$ is
a simple quadratic function of $\sin^{2} \theta $:
\begin{equation}
S(r_{0}, \theta )  =  - \frac{2 \mu \Omega }{r_{0}}
 \sin^{2} \theta 
\left[ 1 - \frac{\sin^{2} \theta }{\sin^{2} \theta_{p} }  
\right],
  \label{eqn.Scal}
\end{equation}%
%%%%
where the coefficient $ B_{0} $
is replaced by the dipole moment 
$ B_{0} =2 \mu /r^{3} _{0} $.

%%%%
%%%
In Fig. \ref{pic:Ssuf2.pdf}, we compare the functions eq.(\ref{eqn.Scal})
and eq.(\ref{eqn.Smono}) truncated at $\theta _{p}$.
The sharp drop of eq.(\ref{eqn.Smono}) at $\theta _{p}$
means that there is a current sheet there. 
Our choice,  (\ref{eqn.Scal}), smoothly goes to zero at $\theta _{p}$, 
but the strength is much smaller.
In order to examine this fact,
we use in the numerical calculation
$\alpha S$
multiplied by a factor $\alpha$.

%%%
Here we summarize the values
at the inner boundary $ r_{0} $: 

\begin{eqnarray}
E_{\theta} &=&-\frac{2 \mu \Omega\cos \theta \sin \theta}{r^2_{0}}, 
\\
E_{\phi} &=&0,
\\
B_{r} &=& -\frac{ 2\mu \cos \theta }{r^3_{0}}, 
\\
%
%B_{\phi} =   - \frac{2 \alpha \mu \Omega }{r^2_{0}}
% \sin \theta 
%\left[ 1 - \frac{\sin^{2} \theta }{\sin^{2} \theta_{p} } 
%\right] .
 B_{\phi} &=& \begin{cases}
      - \frac{2 \alpha \mu \Omega }{r^2_{0}}
 \sin \theta 
\left[ 1 - \frac{\sin^{2} \theta }{\sin^{2} \theta_{p} } 
\right] & (\theta \leq \theta_{p}) \\
    0 & (\theta > \theta_{p}).
  \end{cases}
%.. 
%  \label{eqn.interE}
\end{eqnarray}

On the other hand, the remaining components,
$E_{r}$ and $B_{\theta}, $ cannot be specified.
They are given by radial derivative of the potentials.
%%%%%%%%

%%%
The other boundary conditions are almost clear.
On the pole $\theta =0$,
$E_{\theta}, E_{\phi}$ and $B_{\theta}, B_{\phi}$
should vanish,
such that the functions should satisfy the condition
on the pole
\begin{equation} 
G (r, 0) = 0,
~~
S (r, 0) =0,
~~
F (r, 0) = 0,
~~
\frac{ \partial \Phi (r, 0)}{ \partial \theta} =0.
%
%  \label{eqn.interE}
\end{equation}
%%%
%

We assume a symmetry with respect to the
equator such that
$E_{\theta} =0, B_{\phi } =0$.
On the equator, 
the boundary conditions for 
$S,F$ and $\Phi$
are
\begin{equation}
S(r, \pi/2) =0,
~~
F(r, \pi/2) = 0,
~~
\frac{ \partial \Phi(r, \pi/2)}{ \partial \theta} =0.
%
%  \label{eqn.interE}
\end{equation}
%%%
The conditions for the magnetic flux function $G$ 
are different inside and outside
the light cylinder $\Omega ^{-1}$.
The magnetic flux function $G$ for $r <\Omega ^{-1}$
is 
\begin{equation}
\frac{ \partial G(r, \pi/2)}{ \partial \theta} =0
%G = 0,
\end{equation}
%%%
whereas that for $r \ge\Omega ^{-1}$
\begin{equation}
 G(r, \pi/2) = G_{0} ,
%G = 0
\end{equation}
where $G_{0}$ is a constant describing the last open field line.

Finally, at the outer boundary, 
we impose outgoing condition.
In order to remove numerical reflection at the outer
boundary, the size $r_{out} $ is set to a sufficiently large value, and 
the simulation time is limited to
$ t \le r_{out} $.

\subsection{Electrical conductivity $\sigma$}
The current model eq. (\ref{jmodelbeta0}) has electrical conductivity $\sigma$.
Electrical conductivity $\sigma$ depends on r by the following formula:

\begin{equation}
\sigma (r) = \frac{\sigma_{0}}{r^{n}}. 
\label{sigma0dep}
\end{equation}

Here, $\sigma_{0}$ is the electrical conductivity of the surface $\sigma_{0} = \sigma (r=1)$;
$n$ is r-dependence parameter.
The electrical conductivity is large in the vicinity of the star and  falls off with increasing radial distance.
The outer boundary is set to an almost-vacuum condition.
We now consider each of the electrical-conductivity parameters $\sigma_{0} =2,5,10,20,50,100 $ in the $n=2$ case.

\section{Result}%4.3
\subsection{The structure of the obtained solution}%4.3
The dynamical equations (\ref{eqn.G}), (\ref{eqn.S}), (\ref{eqn.P}), and (\ref{jmodelbeta0}) are integrated by time, and almost stationary solutions are obtained after several wave-crossing periods.
Steady states appeared under each electrical conductivity parameter $\sigma_{0} =2,5,10,20,50,100 $ in the $n=2$ case.
The results for $n=2$ and $\sigma_{0}=10,50$ are explained in the next section.

\subsubsection{Poloidal magnetic field, G}
\begin{figure*}
\gridline{\fig{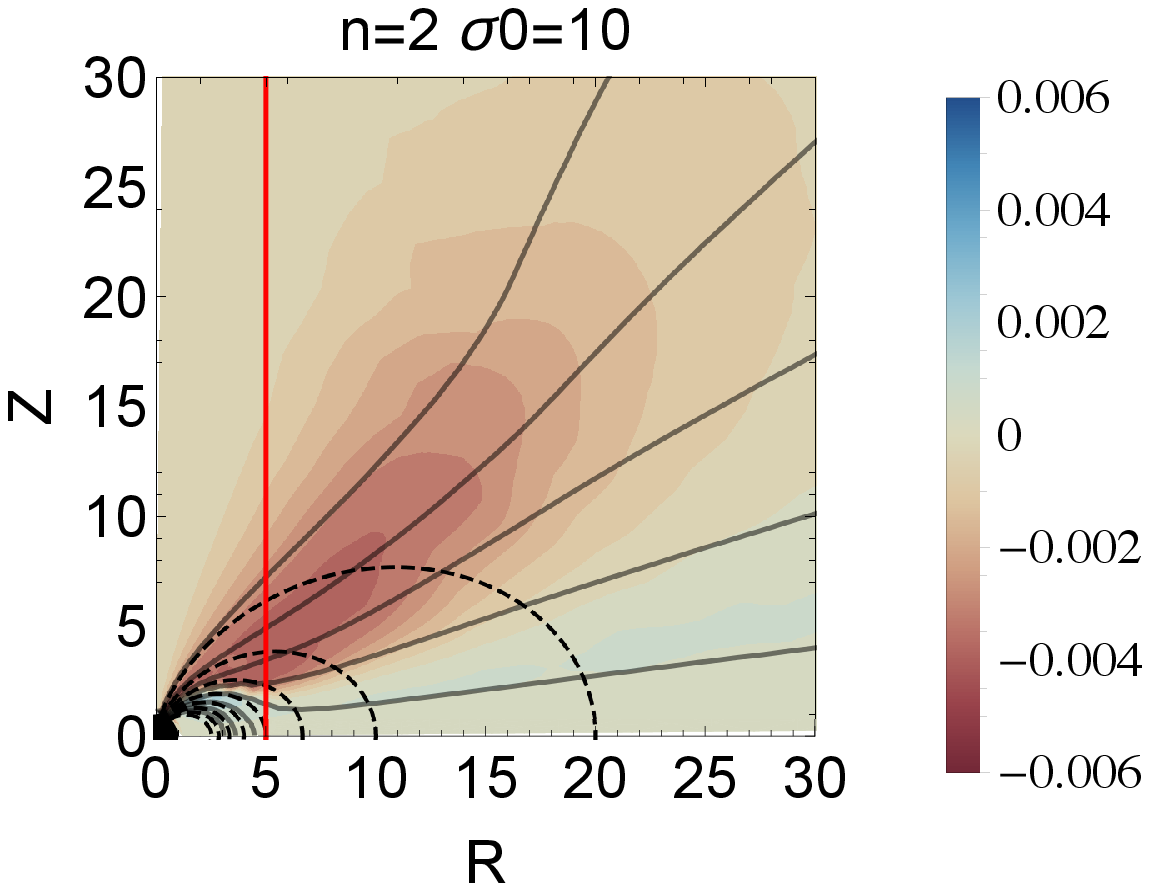}{0.5\textwidth}{(a)}
          \fig{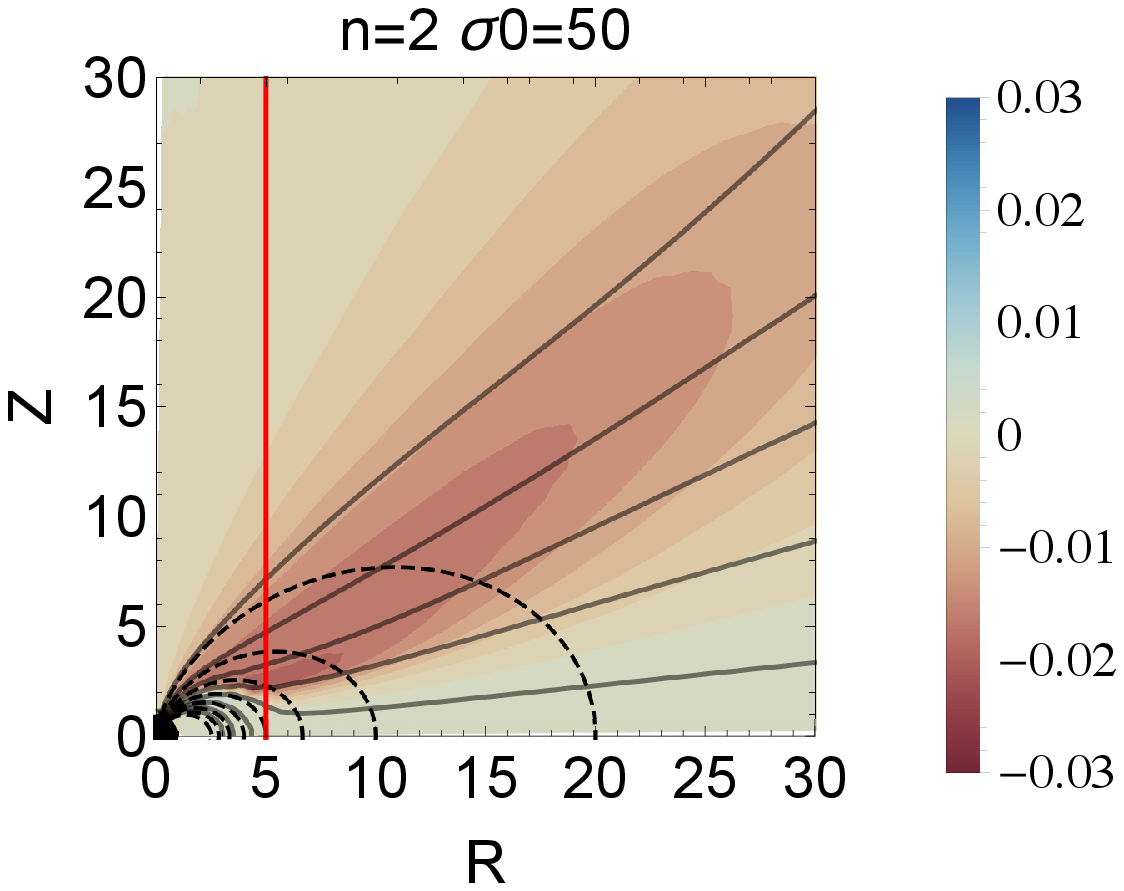}{0.5\textwidth}{(c)}}
\gridline{\fig{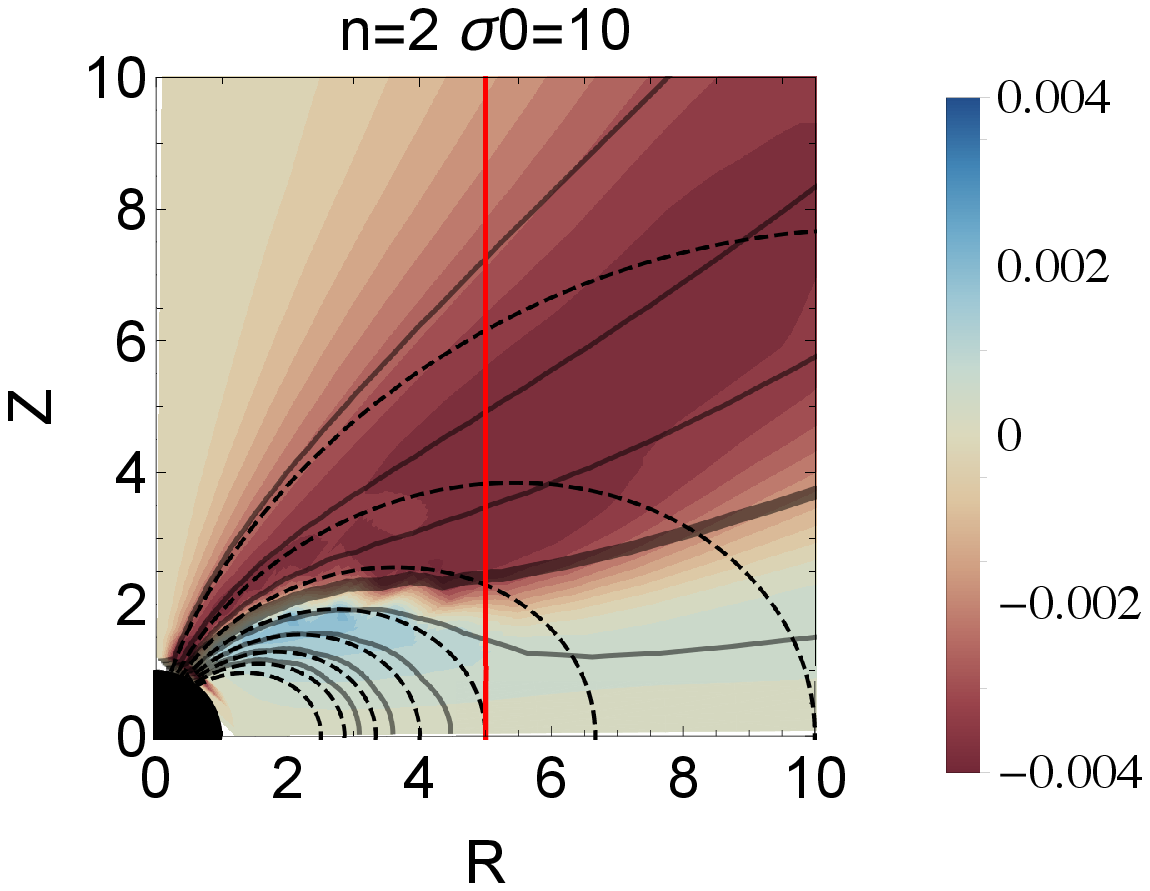}{0.5\textwidth}{(b)}
\fig{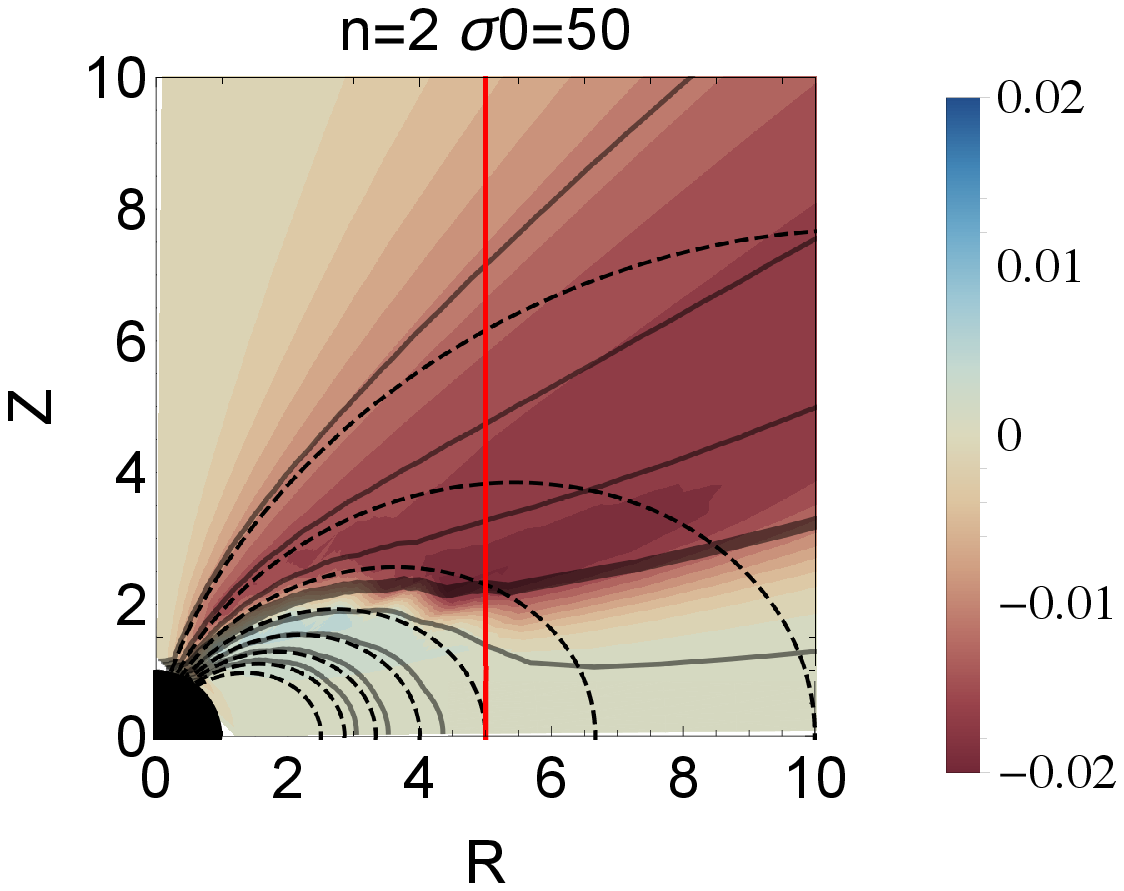}{0.5\textwidth}{(d)}}
\caption{Poloidal and  toroidal magnetic fields  in the meridian plane for $\sigma_{0}=10,50$. The light cylinder radius $R_{LC} = c/\Omega $ is shown by a red vertical line. The poloidal magnetic field line is solid and black. The dashed black lines show the magnetic dipole poloidal magnetic fields. The toroidal magnetic strength is indicated by contour color. Bottom is zoom in view.}
\label{pic:n2n10st200.pdf}%a
\label{pic:n2n50st200.pdf}%c
\label{pic:n2sig10GS200.pdf}%b
\label{pic:n2sig50GS200.pdf}%d

\end{figure*}

The dynamical equations are integrated by time, and almost-stationary solutions are obtained after several wave-crossing periods.
From the star surface to the outside area, the magnetosphere gradually becomes a steady state.
Specifically, the magnetic field at radius $r=30$ became steady over the time period $t=50$.
I explain the formation of the magnetic field in terms of its poloidal and toroidal components separately.
Figure \ref{pic:n2n10st200.pdf}(a) shows the steady state magnetic fields  $n=2, \sigma_{0}=10$ in the meridian plane. 
The red line in Fig. \ref{pic:n2n10st200.pdf}(a) is a light cylinder of radius $R=R_{LC}=5$.
The black solid lines represent the flux functions $G/(\mu/r_{0}) =0.05,0.1,...0.4$, normalized by the dipole poloidal magnetic flux in surface.
In particular, the thick black line of Fig \ref{pic:n2sig10GS200.pdf}(b) is $G/(\mu/r_{0}) =0.2$.
The magnetic field lines, $G=0.2$, pass through the light cylinder in the case of the magnetic dipoles.  
For comparison, those for the magnetic dipole are also shown by black dashed lines with the same values, $G_d/(\mu/r_{0}) =0.05,0.1,...0.4$.
The lines with $G/(\mu/r_{0}) <0.285$ are open outwardly, and the others are closed  within the light cylinder.
Both lines for the dipolar and numerical results of, e.g., $G/(\mu/r_{0}) =0.05$, are parallel for $r/r_{0} < 3$, but they are significantly separated beyond the light cylinder.
The line for the vacuum dipole is closed, whereas that for our numerical result is open, and the structure becomes almost radial. 
Open and closed magnetic-field lines feature the same force-free-electrodynamics pulsar solution.
The magnetic field is very strong and it moves with the plasma; it cannot co-rotate because the co-rotation speed outside of the light cylinder exceeds the speed of light. 
Thus, the magnetic-field lines are open for the outside.
Specifically, $G=0.285$ in the equatorial plane is greater than 0.2, toroidal current  form a poloidal magnetic fields in the vicinity last open field line.
Next, I show the result of changing the electrical-conductivity parameter $\sigma_{0}$.
Fig. \ref{pic:n2n10st200.pdf}(a) and Fig.\ref{pic:n2n50st200.pdf}(c) respectively correspond to the electrical conductivities $\sigma_{0}=10$ and $50$.
Comparison of these two figures show that they exhibit poloidal magnetic fields of the same shape.

%%%%%%%%%%%%%%%%%%%%%%%% %%%%%%%%%%%%%%%%%%%%%%%%%%%%%%%%
\subsubsection{Toroidal magnetic field S}

The toroidal magnetic field $ B_{\phi}=S/R$ is also shown in Fig. \ref{pic:n2n10st200.pdf}(a).
The color contour represents $S/(\mu /r_{0}^2)$.
Red and blue indicate negative and positive values, respectively.
The maximum value of $S$ is located inside the light cylinder.

Fig. \ref{pic:n2sig10GS200.pdf}(b) shows the magnetic field near the star surface.
According to Figure \ref{pic:n2sig10GS200.pdf}(b), the toroidal field S has a negative value in  the $G<0.2$ region and a positive value in the $G>0.2$ area near the equatorial plane. The poloidal current inwardly flows in the polar region, i.e., the region from the z-axis to the maximum with respect to $\theta$, whereas the current outwardly flows along the last open line in the equatorial region.  
The toroidal magnetic field $S$ gradually goes to zero outwardly due to resistivity and spreads to the outside beyond the light cylinder.
No toroidal magnetic field, S, exists near the pole. 
This result is due to spatial symmetry. 

\begin{figure}
\begin{center}
\includegraphics[angle=0,trim=1cm 0cm 0.5cm 0.5cm,width=1.02\columnwidth]{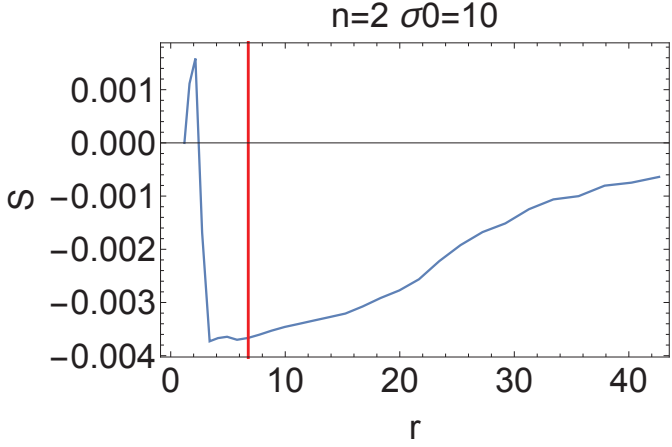}
\caption{Time-averaged toroidal magnetic flux function S for the center of the star. $\theta=0.74$}
\label{pic:timeav_s_sig10.pdf}
\end{center}
\end{figure}

Fig. \ref{pic:timeav_s_sig10.pdf} shows the time averaged toroidal flux S in $\theta=0.74$.
The toroidal magnetic field S decreases gradually by the radius from the star.
The maximum value of S falls inside the light cylinder radius.
The maximum value does not match the radius of the light cylinder.
The concentrated poloidal current  area is the light cylinder interior.

The toroidal magnetic field, S, increases due to large electrical conductivity $\sigma_{0}$.
In our model, the value increases with the increase of $r$ eq. (\ref{sigma0dep}).
The toroidal magnetic field increase proportionally to electrical conductivity.
Fig.\ref{pic:n2n50st200.pdf}(c) shows the electromagnetic field for $\sigma_{0} = 50$, five times bigger than the $\sigma_{0}=10$ case. 
In addition, this field area spreads out.
This is in contrast to the poloidal magnetic field G, which is not changed by the electrical conductivity.

\subsubsection{Electrostatic potential $\Phi$}

\begin{figure*}
\gridline{\fig{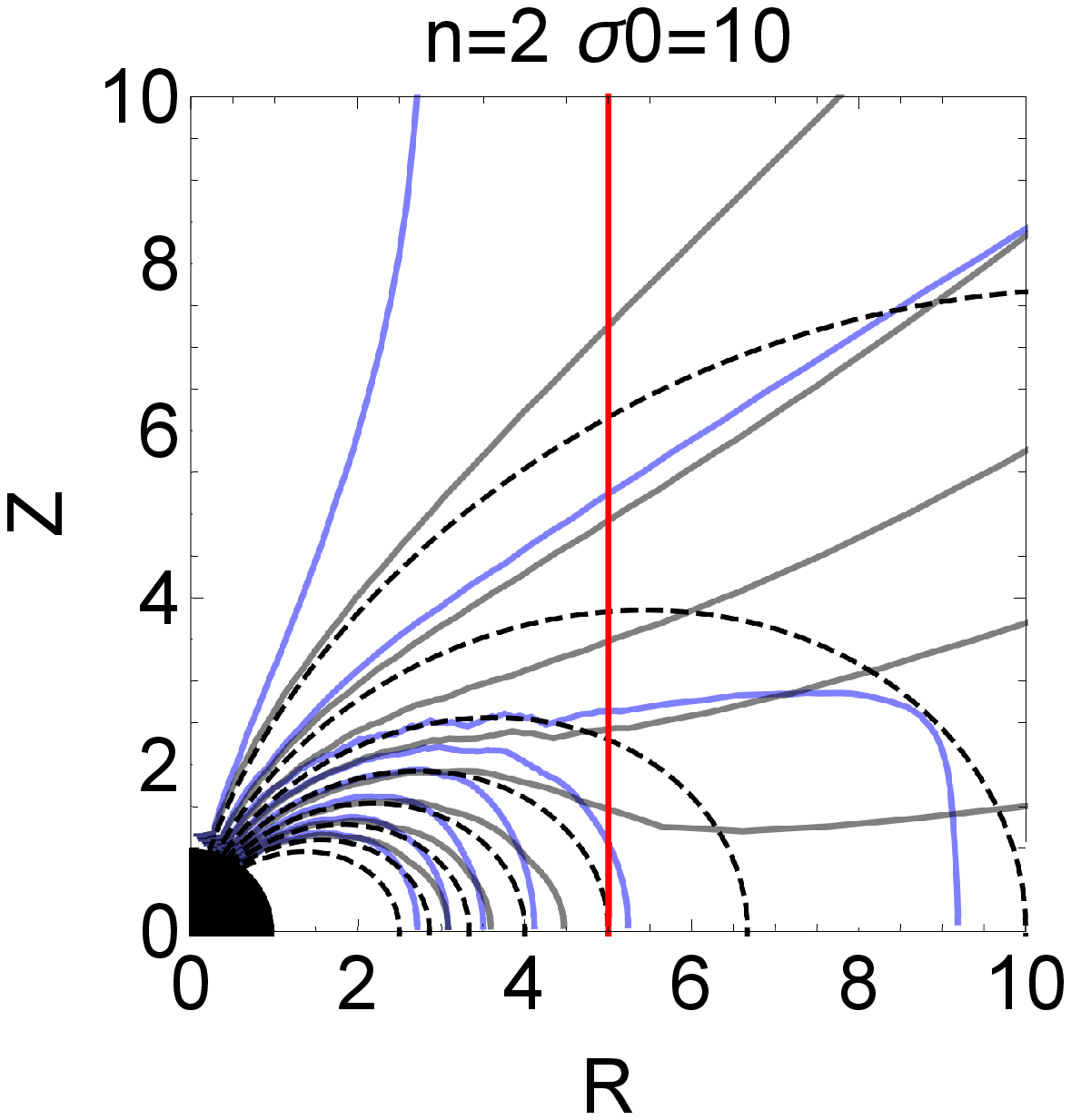}{0.5\textwidth}{(a)}
          \fig{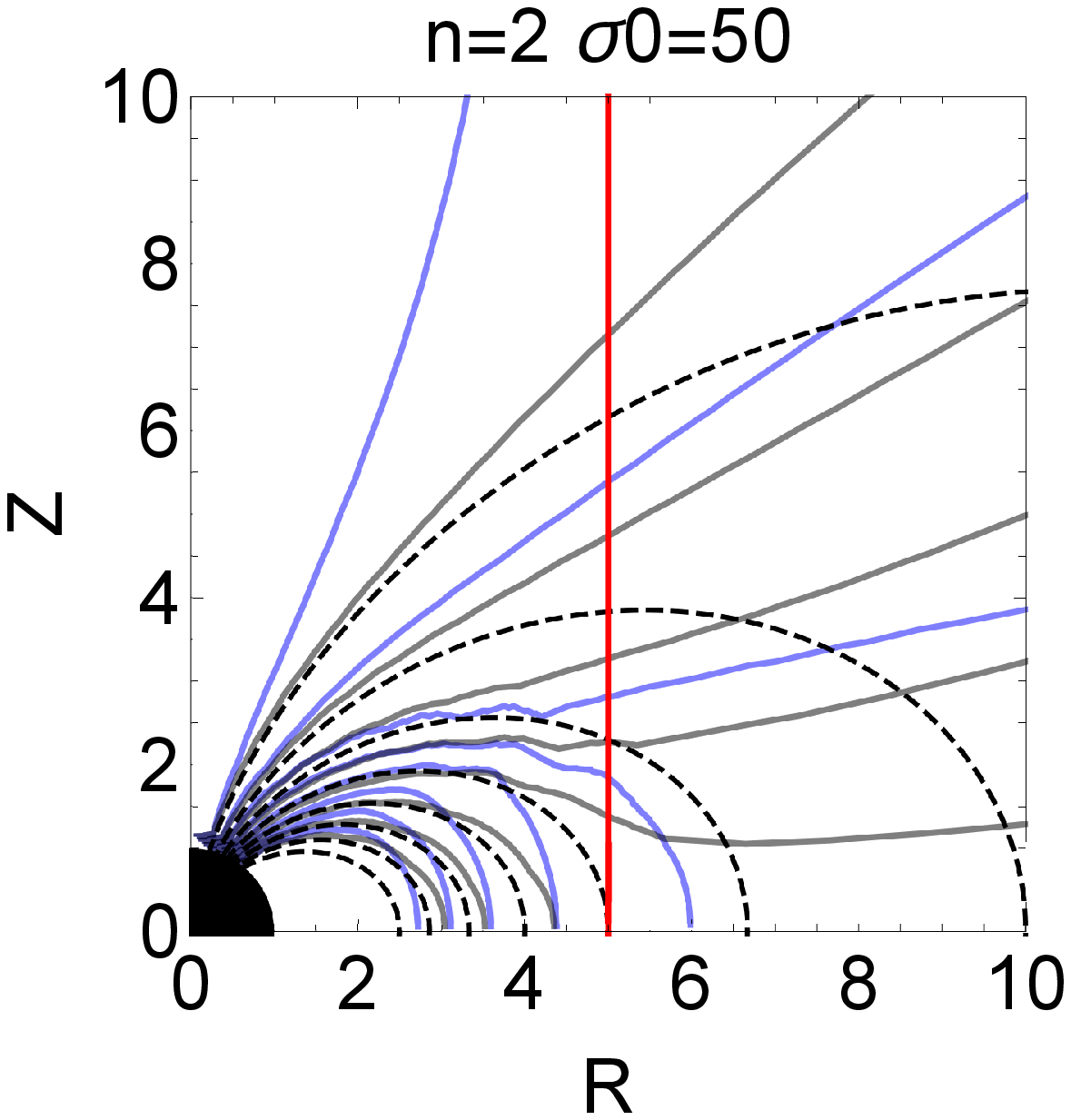}{0.5\textwidth}{(c)}}
\gridline{\fig{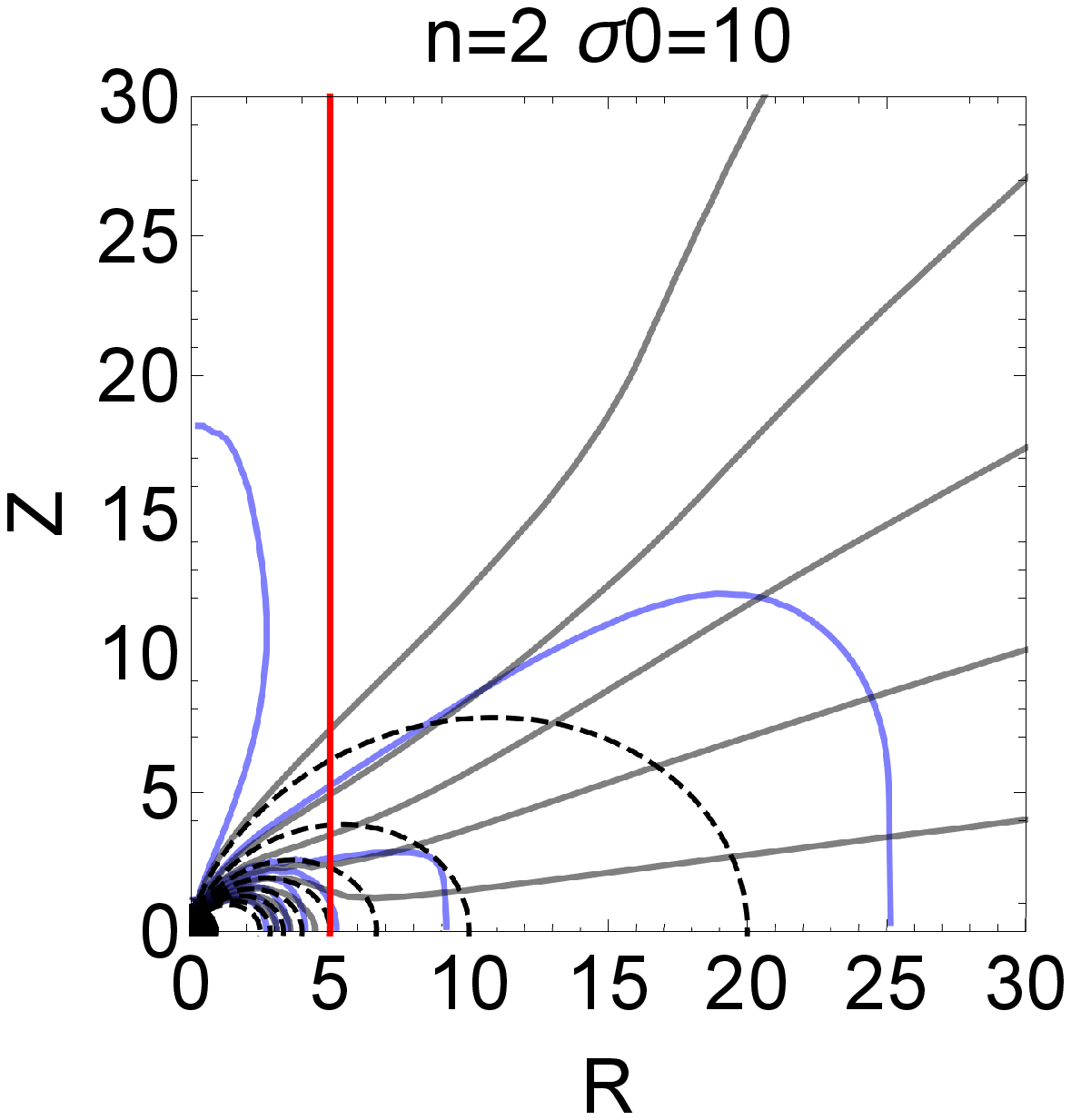}{0.5\textwidth}{(b)}
\fig{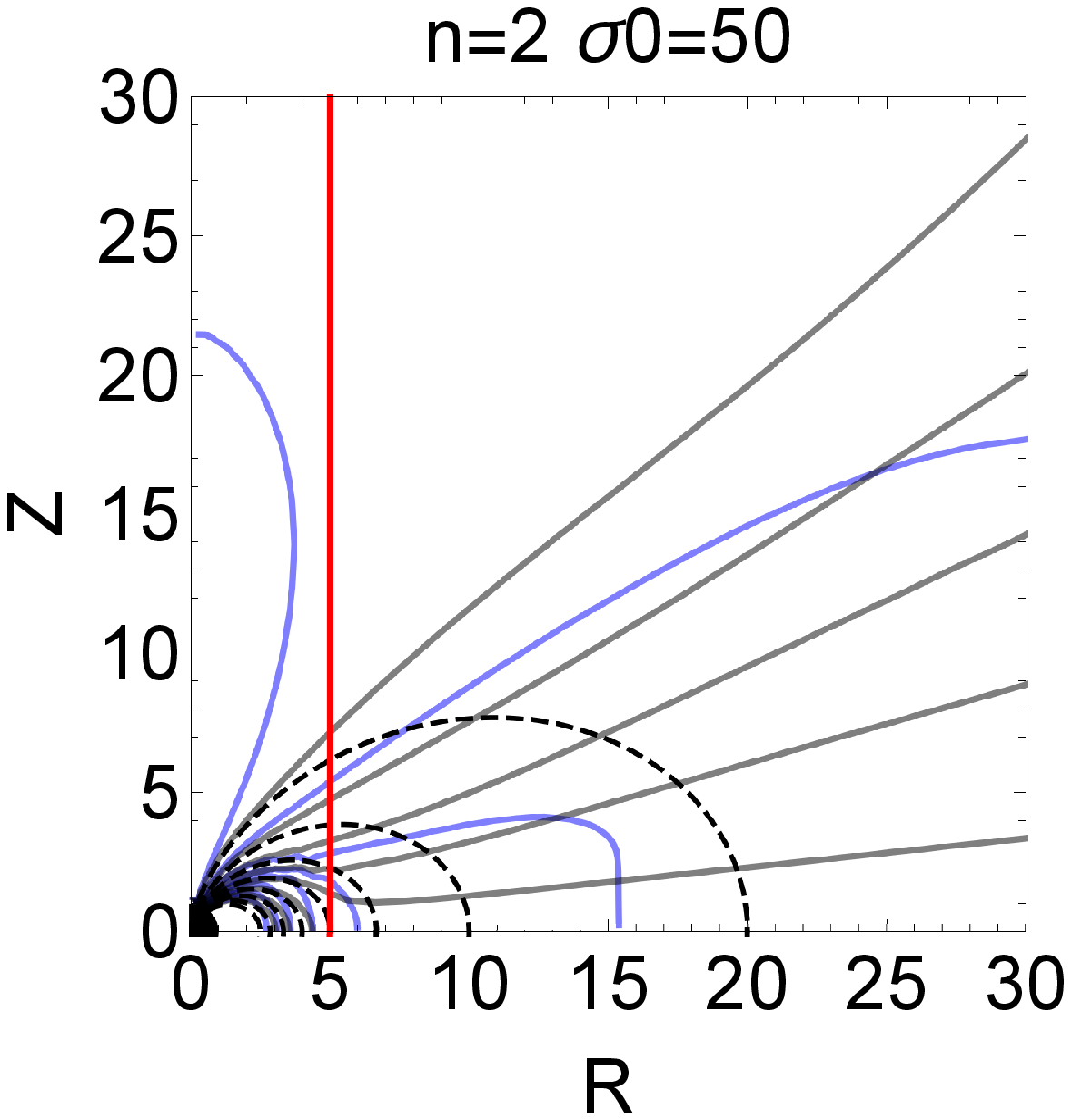}{0.5\textwidth}{(d)}}
\caption{Electrical potential $\Phi$ for $n=2,\sigma_{0}=10,50$. The red line is the light cylinder radius $R=5$. The black line is the equal value of the poloidal-magnetic flux function $G$. The black dotted line shows a dipole magnetic flux  equal to the value of the function. Bottom is zoom out view.}
\label{pic:n2sigma10Gphi200.pdf}%a
\label{pic:n2sigma50Gphi200.pdf}%c
\label{pic:n2sigma10GPhi200soto.pdf}%b
\label{pic:n2sima50GPhi200soto.pdf}%d
\end{figure*}

Fig. \ref{pic:n2sigma10Gphi200.pdf}(a) shows the electrostatic potential contour of $\sigma_{0} = 10$.
From $ \Phi = 0.01 $ to 0.08, contours are equally spaced for each 0.01.
In the numerical calculation, the electrostatic potential on the star's surface does not change with time because of the boundary conditions eq. (\ref{boundphi}).
Blue lines in Fig. \ref{pic:n2sigma10Gphi200.pdf}(a) spread out from the surface boundary and extend in the radial direction.
Blue lines are connected vertically towards the equatorial plane because the field's boundary condition has only an R component in this plane.
Fig. \ref{pic:n2sigma10GPhi200soto.pdf}(b) shows the electrostatic potential of the spatial region from $Z=0$ to $30$.
Here, we have drawn the contour line in increments of 0.005 from $\Phi = 0.04$.

The densities of the equipotential lines are proportional to the strength of the electric field, which is strong on the star surface and decreases as one goes radially outward.
No electric field dominant ($ E^{2} -B^{2}> 0 $) area  exist in $ 1 \leq r \leq 30$.
The magnetic field in the obtained results is always stronger than the electric field.

Fig.\ref{pic:n2sigma50Gphi200.pdf}(c) is the drawn contours of the $\sigma_ {0} = 50$ electrostatic potential cases.
Electrical conductivity $ \sigma_ {0} $ increases along with the electrostatic potential of the equatorial plane.
However, the difference of electrostatic potential of the equatorial plane between the $\sigma_{0} = 10$ and $\sigma_{0} = 50$ is about 10\%.

When the electrical conductivity is different from this case, the overall structure of the solution remains similar. 
For example, electrical conductivity is increased in a toroidal magnetic field, but little shape change 
was observed about the potential.

%%%%%%%%%%%%%%%%%%%%%%%%%%%%%%%%%%%%%%%%%%%%%%%%%%%%%%%%%%%%%%%%%%%%%%%%%%
\subsubsection{Poloidal current $j_{p}$}
%%%%%%%%%%%%%%%%%%%%%%%%%%%%%%%%%%%%%%%%%%%%%%%%%%%%%%

Current density is divided into poloidal and toroidal components.
The poloidal component of  current density $j_{p}$ consists of $j_{r}$ and $j_{\theta}$.

\begin{eqnarray}
j_{p} = \sqrt{ j_{r}^{2} + j_{\theta}^{2}}.
\end{eqnarray}

\begin{figure*}
\gridline{\fig{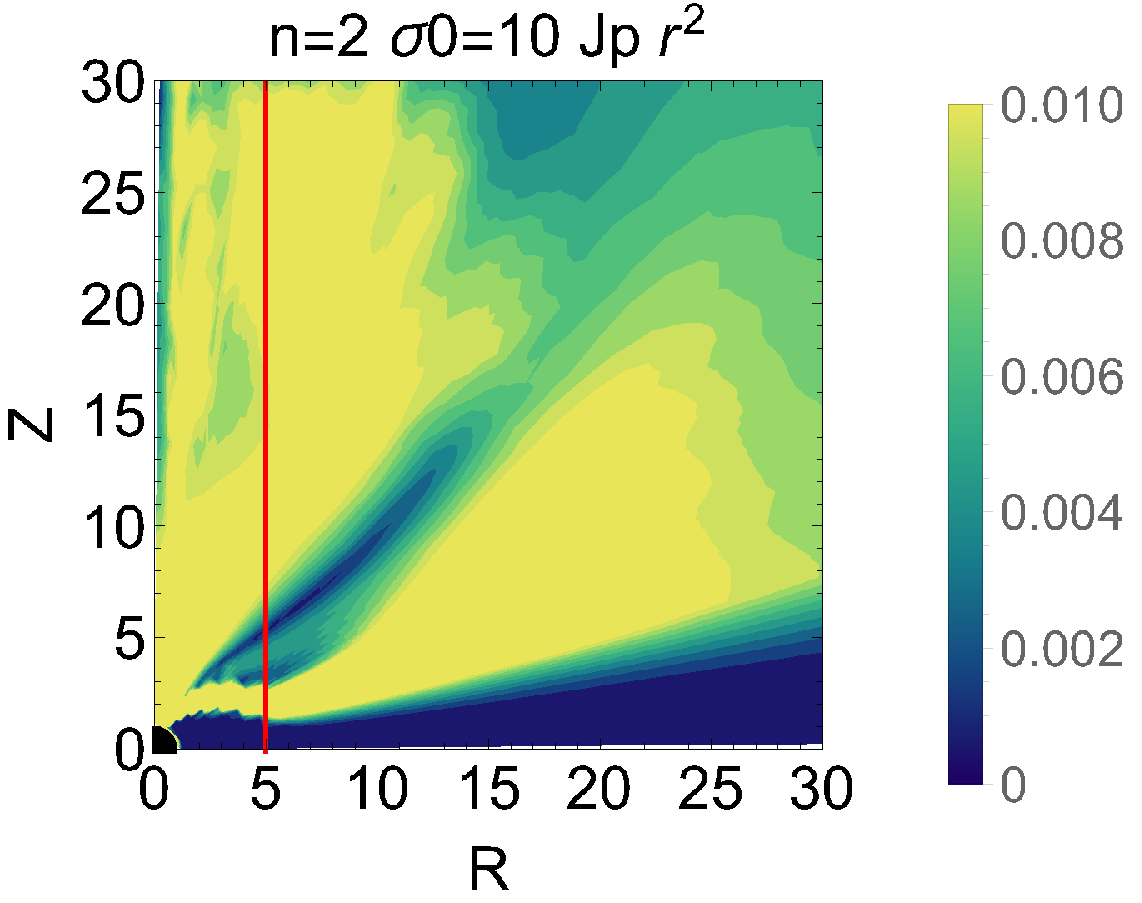}{0.45\textwidth}{(a)}
          \fig{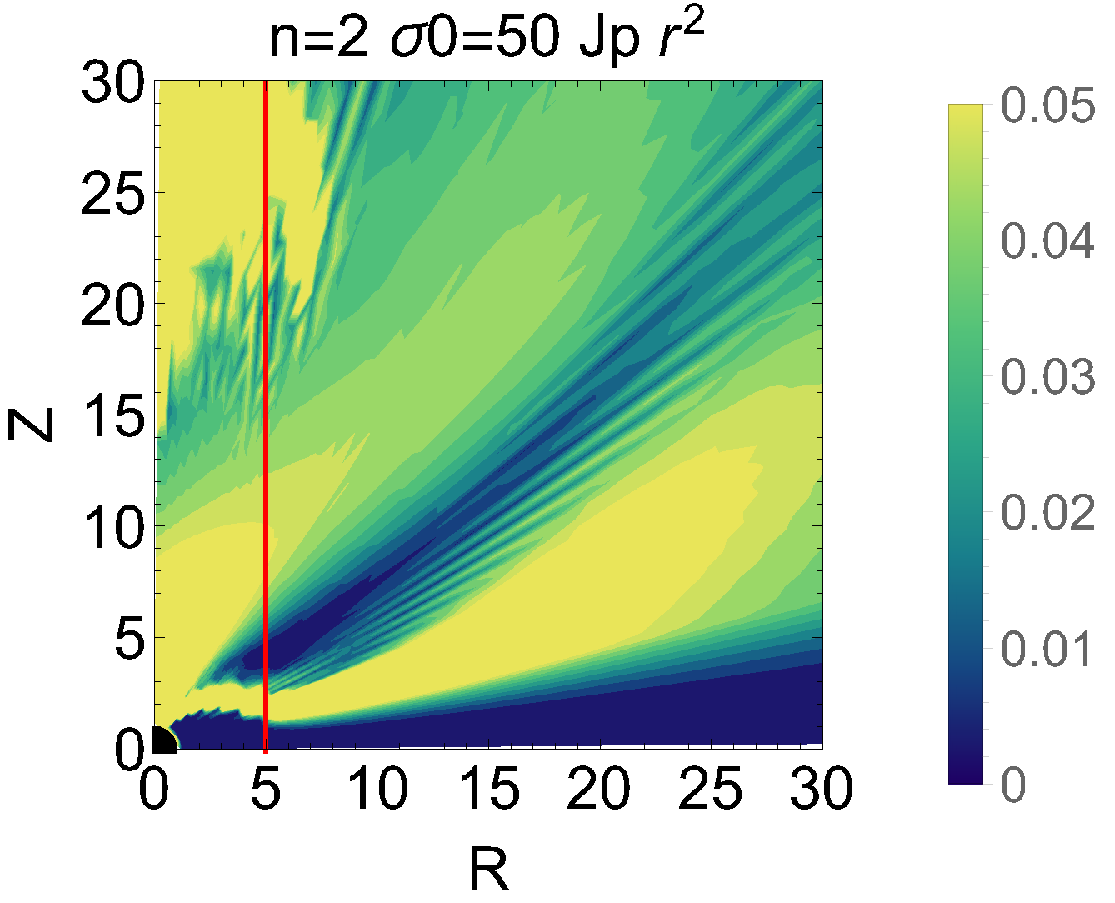}{0.45\textwidth}{(d)}}
\gridline{\fig{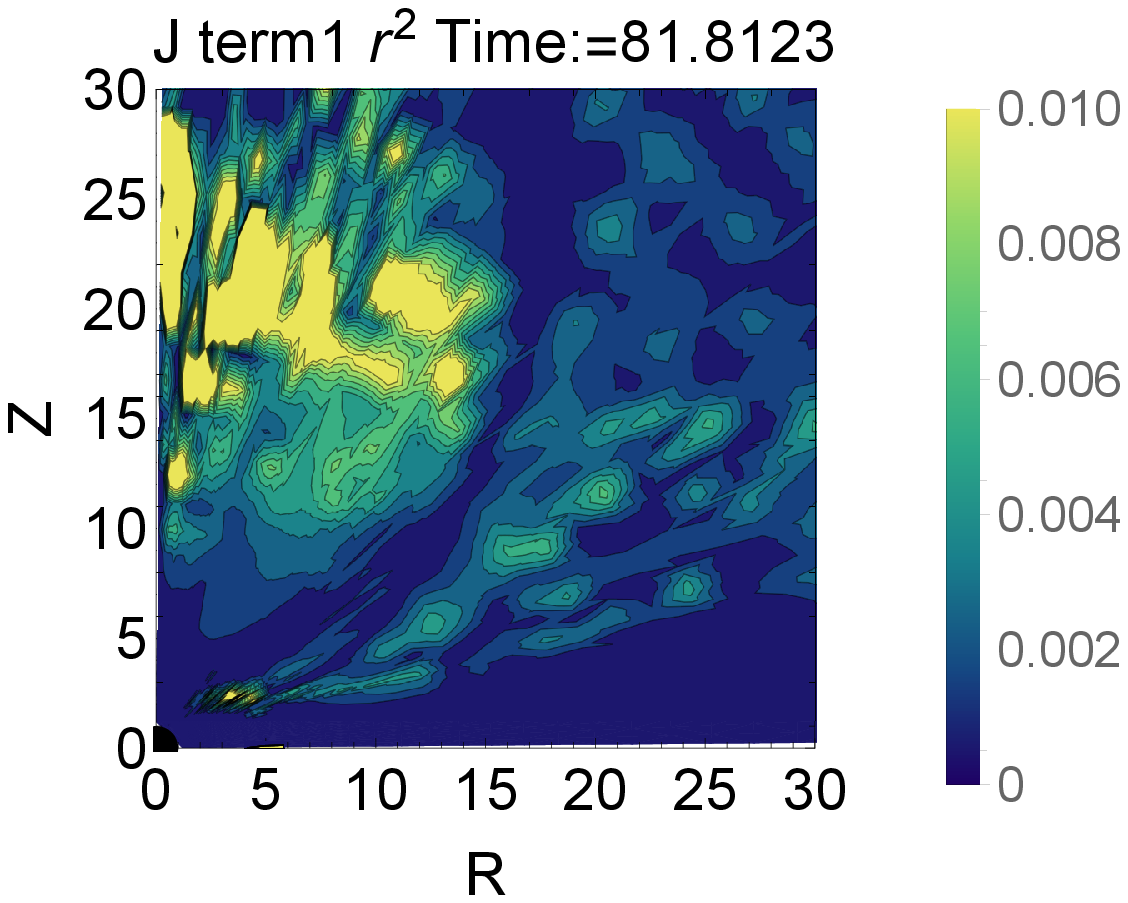}{0.45\textwidth}{(b)}
\fig{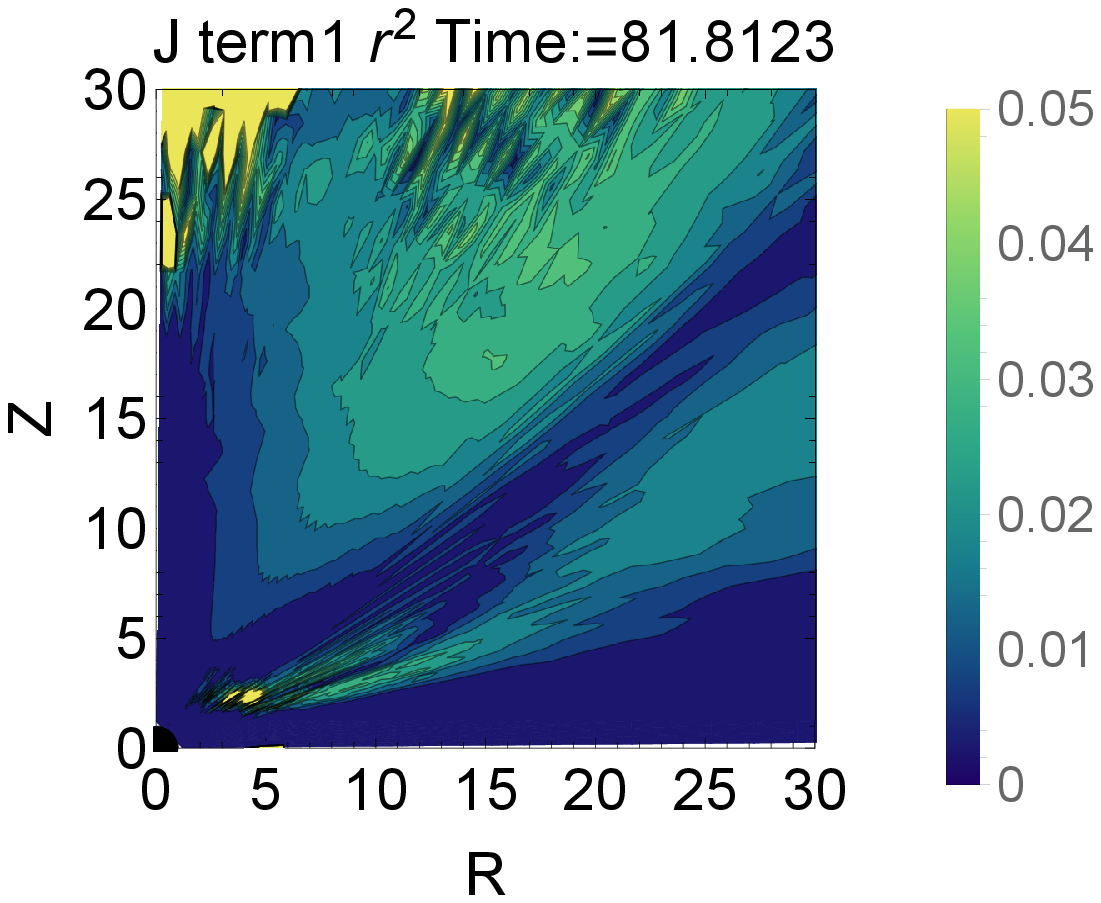}{0.45\textwidth}{(e)}}
\gridline{\fig{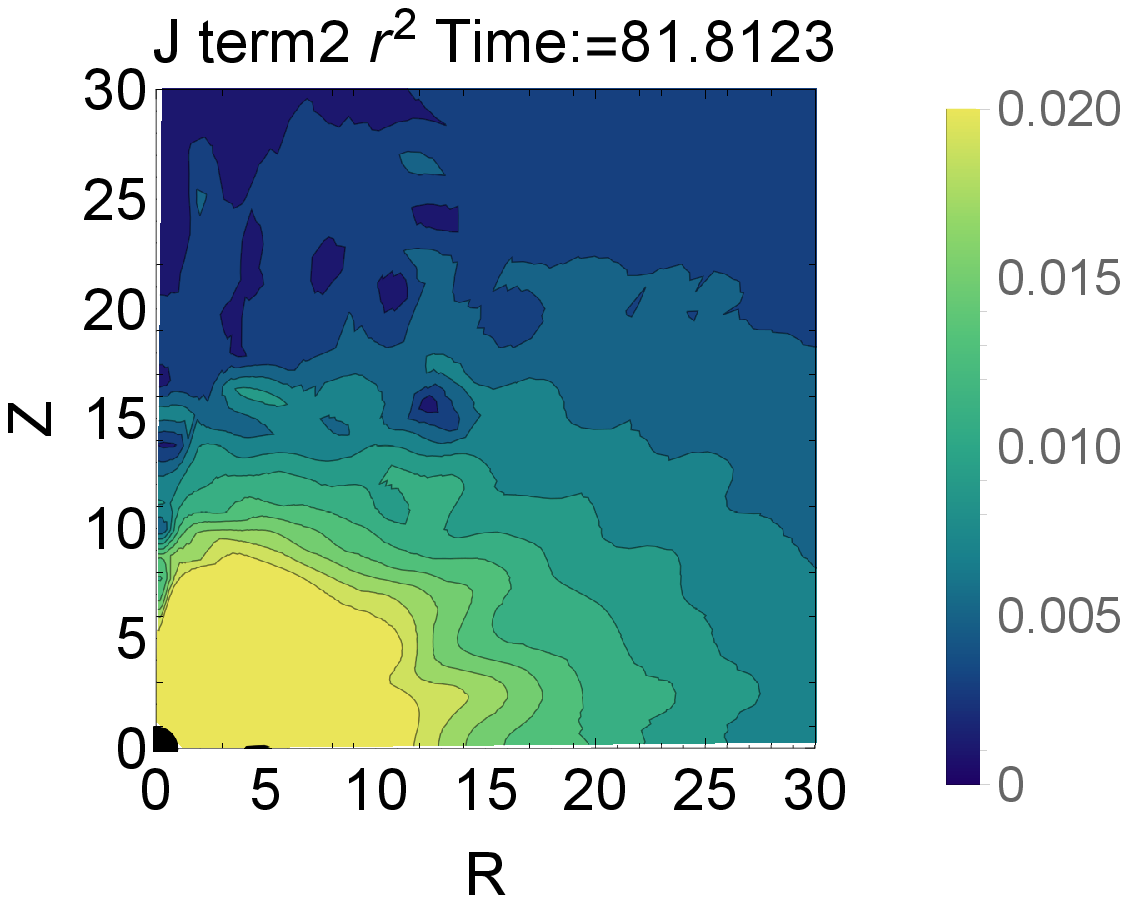}{0.45\textwidth}{(c)}
          \fig{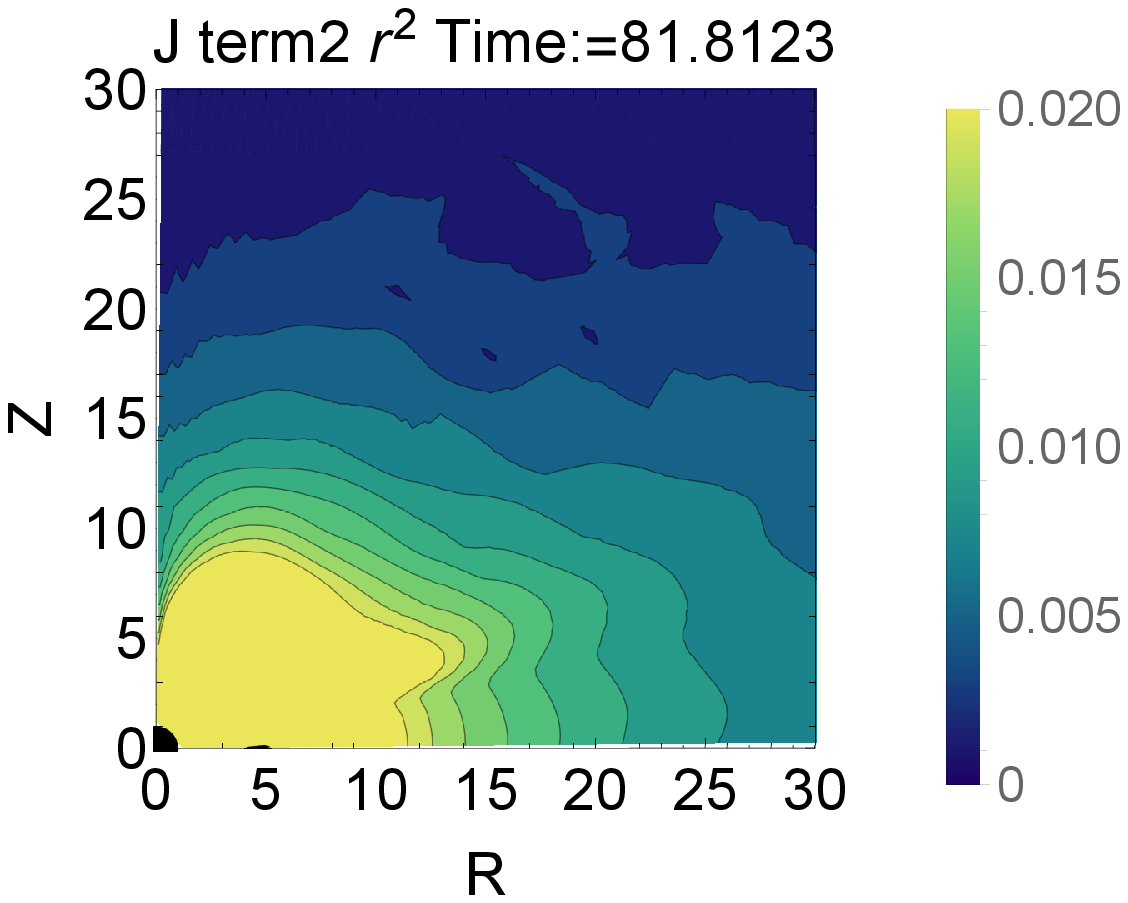}{0.45\textwidth}{(f)}}
\caption{Top:Poloidal current density $j_{\mathrm{p}} r^{2}$ for $n=2,\sigma_{0} =10,50$. Middle:Poloidal current density of first term component in  eq.(\ref{jmodelbeta0}) for $\sigma_{0}=10,50$. Bottom:Poloidal current density of second term component in  eq.(\ref{jmodelbeta0}) for $\sigma_{0}=10,50$.}
\label{pic:N2N10_jprS1_00090.pdf}%a
\label{pic:N2N50_jprS1_00090.pdf}%d
\label{pic:jterm1n2n10s.pdf}%b
\label{pic:jterm1n2n50s.pdf}%e
\label{pic:jterm2n2n10.pdf}%c
\label{pic:jterm2n2n50.pdf}%f
\end{figure*}

Fig. \ref{pic:N2N10_jprS1_00090.pdf}(a) plots the magnitude of the poloidal current density at electric conductivity $\sigma_{0}=10$.
The yellow region has a large current density; the blue region has a small current density.
Fig. \ref{pic:N2N10_jprS1_00090.pdf}(a) shows that an electric circuit forms in the magnetosphere.
Fig. \ref{pic:N2N50_jprS1_00090.pdf}(d) shows the poloidal current density  $j_{p} r^{2}$ for $\sigma_{0}=50$. 
Both figures exhibit the same poloidal current circuit structure.
I divide three parts  to understand the mechanics of the poloidal current density distribution.  
Fig. \ref{pic:kairo.pdf} shows a current circuit schematic diagram.
The oblique lines in the central region indicate low poloidal current density.
The arrows A and B are the current inside of the light cylinder; both parts of current flow along the magnetic field line but in opposite directions.
Arrow C indicates current that does not flow along the magnetic field, but returns inside.
The total current flow composes the current circuit and gives rise to the toroidal magnetic field.
Poloidal current makes the toroidal magnetic field.
The oblique lines in the region near the star indicate a dead zone.
In this region, plasma co-rotates with the star; thus, the poloidal current does not exist.

The r-component electric field $E_{r}$ is formed on the star surface.
Therefore, current flows to the inside of $\theta_{\mathrm{pc}}$ from the star surface.
The poloidal current density, $j_{r}$, at $\theta_{\mathrm{neutral}}$ is zero.
The $\theta_{\mathrm{neutral}}$ star surface represents a boundary at which the direction of the poloidal current is interchanged.
Where $0 < \theta < \theta_{\mathrm{neutral}}$ (Arrow A), $j_{r}<0$ because there is a current toward the star.
In addition, there is a current flowing outward from the star, $\theta_{\mathrm{neutral}}< \theta <\theta_{\mathrm{p}}$ (Arrow B).
The region $\theta > \theta_{\mathrm{p}}$ is a closed zone in which poloidal current does not flow, meaning that there is a region of low poloidal density besides the star's surface.
The concentration of poloidal current caused by the Lorentz force.
Concentration of the current occurs on the upper and lower sides (Arrow A and Arrow B).
The poloidal current density of the open magnetic field lines for the area outside of the light cylinder also spread spatially.
Focusing on large areas of lower current density (Arrow C), and has a distribution that across the magnetic field lines.
This corresponds to a large Z component of the current vector.
In order to clarify, I compare the magnitudes of poloidal current in different component sections.

\begin{figure}
\begin{center}
\includegraphics[angle=0,trim=1cm 0cm 0.5cm 0.5cm,width=0.95\columnwidth]{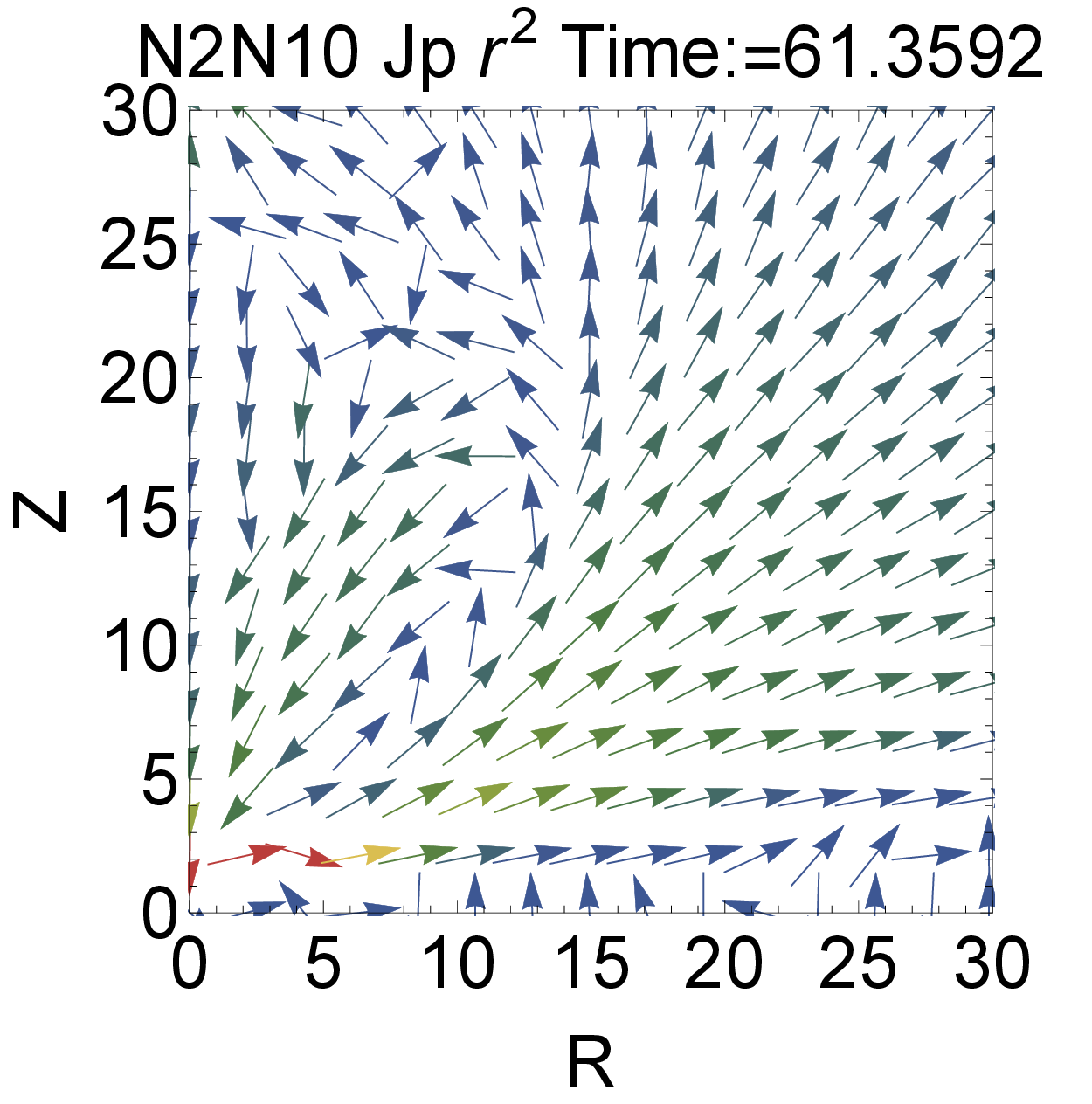}
\caption{For $n=2,\sigma_{0}=10$. Poloidal current vector direction shown in the meridional plane. The color of the vector indicates the intensity of poloidal current from blue to red.}
\label{pic:n2n10den150_3.pdf}
\end{center}
\end{figure}

\begin{figure}
\begin{center}
\includegraphics[angle=0,trim=1cm 0cm 0.5cm 0.5cm,width=0.95\columnwidth]{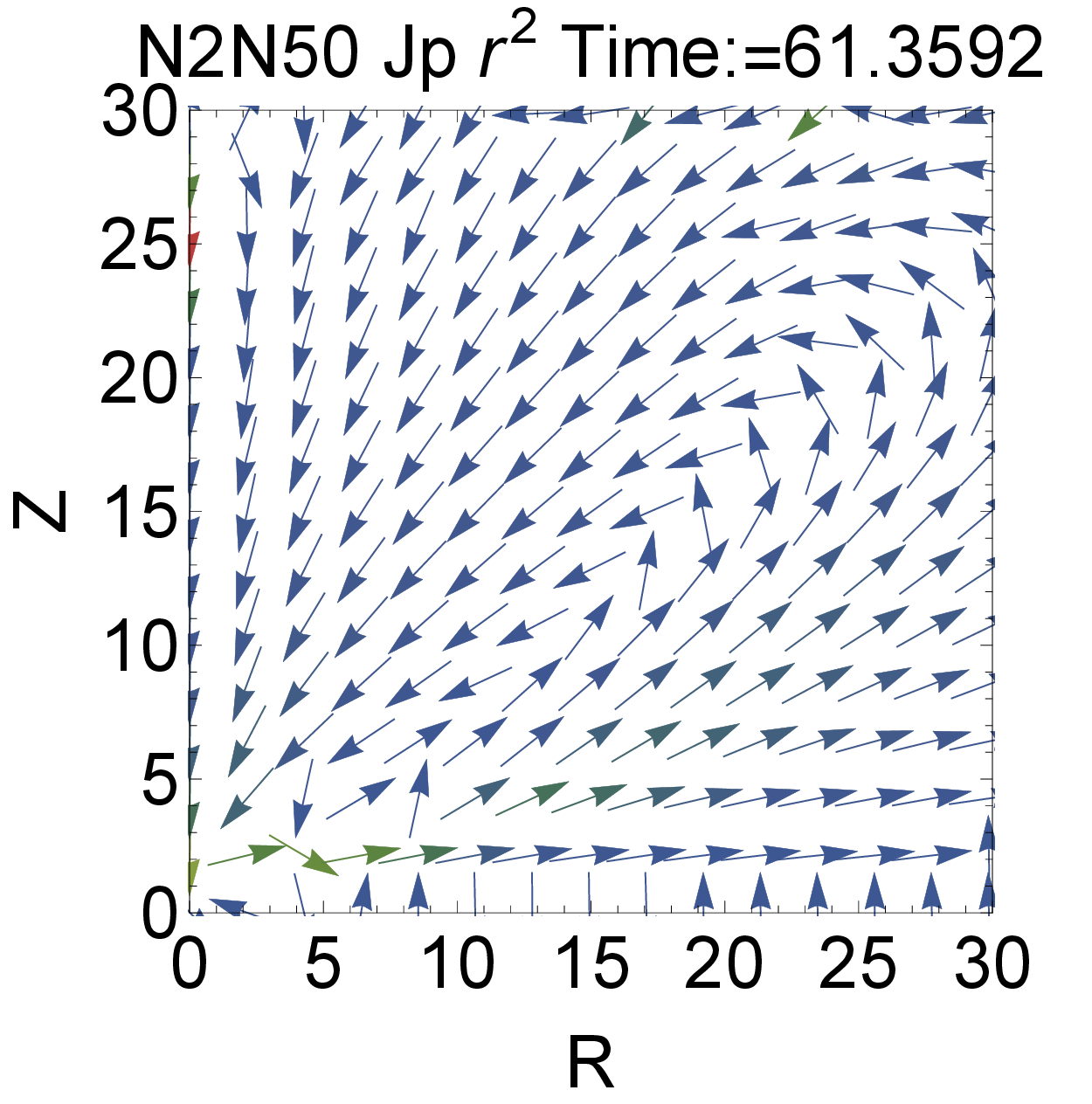}
\caption{Same as Fig.\ref{pic:n2n10den150_3.pdf} for $\sigma_{0}=50$.}
\label{pic:n2n50den150_3.pdf}
\end{center}
\end{figure}

\begin{figure}
\begin{center}
\includegraphics[angle=0,trim=1cm 0cm 0.5cm 0.5cm,width=0.95\columnwidth]{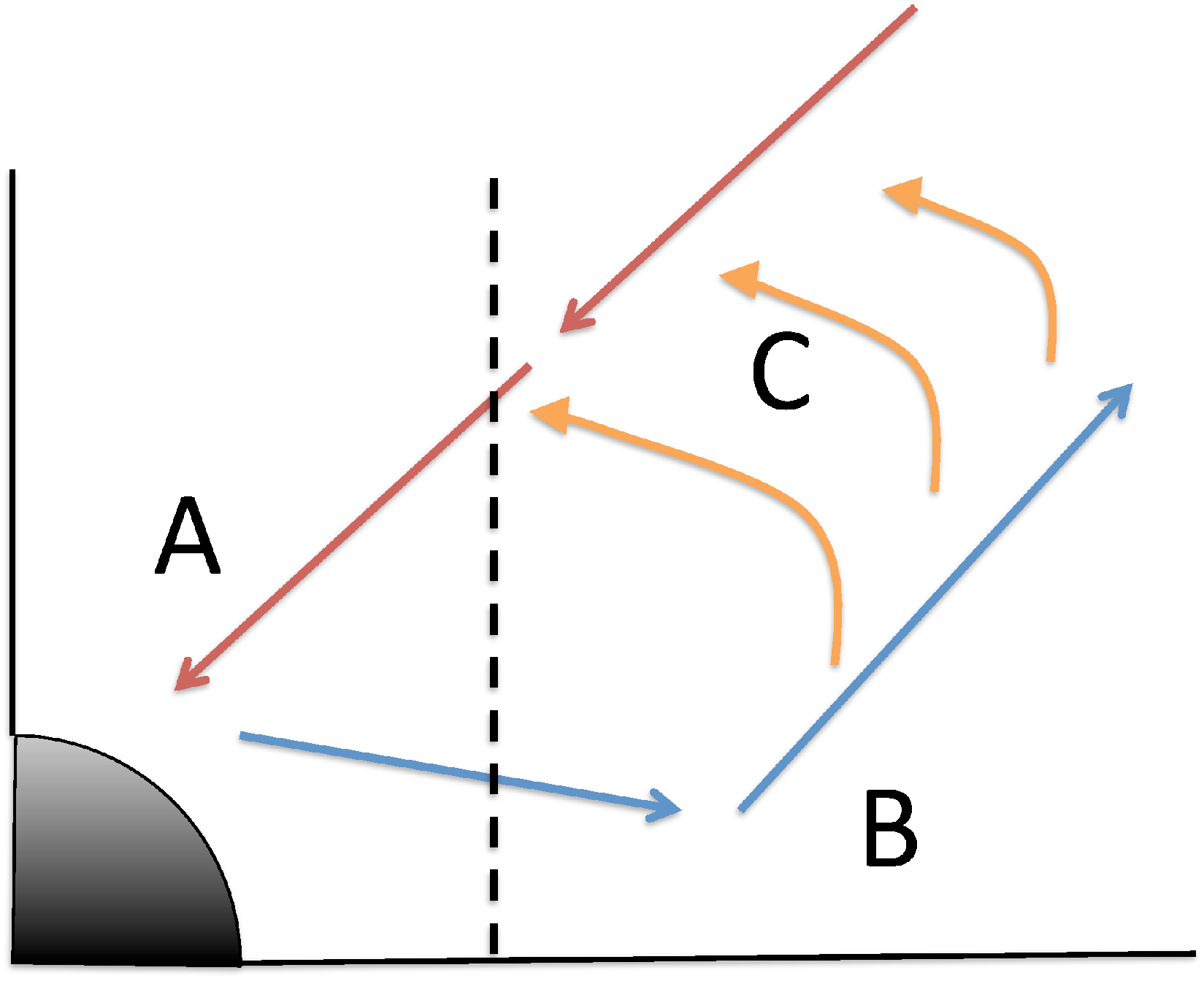}
\caption{This diagram shows poloidal current circuit. The A and B arrows are the current flow along with the magnetic field. The C arrow shows the current perpendicular to the magnetic-field line. This current circuit shows the toroidal magnetic field.}
\label{pic:kairo.pdf}
\end{center}
\end{figure}

Figs. \ref{pic:jterm1n2n10s.pdf}(b) and (c) are obtained by drawing a distribution of the first term and the absolute value of the second term of the poloidal current equation (\ref{jmodelbeta0}). 
The $B_{\phi}E_{r}$ component of the first term becomes dominant due to the fact that the electrical conductivity $\sigma$ decreases outward, making the second term small.
The current along the magnetic field hardly flows to the outside because it is assumed that the electrical conductivity $\sigma$ has a spatial dependence.
Also, the electric field has a component perpendicular to the magnetic field.
The poloidal current perpendicular to the magnetic field is generated by the electric field.
Therefore, the flow of current toward the star across the magnetic-field lines occurs gradually. When this flow returns to the star, one round of the current circuit is formed.
Figs. \ref{pic:n2n10den150_3.pdf} and \ref{pic:n2n50den150_3.pdf} are obtained by drawing the direction of the vector of the poloidal currents. The strength of the poloidal current is increased toward the red from the blue.

Fig. \ref{pic:N2N50_jprS1_00090.pdf}(d) illustrates the poloidal current density at electric conductivity $\sigma_{0}=50$.
In the case of $\sigma_{0}=20$, the $j_{p}=0$ area is eliminated by $r=10$.
In the case of $\sigma_{0}=50$, the poloidal current density distribution is changed, as shown in \ref{pic:N2N50_jprS1_00090.pdf}(d);
this is because the components perpendicular to the magnetic lines of force become dominant.

Figs. \ref{pic:jterm1n2n50s.pdf}(e) and (f) are each of the poloidal current densities in the cases where $\sigma_{0}=50$.

%%%%%%%%%%%%%%%%%%%%%%%%%%%%%%%%%%%%%%%%%%%%%%%%%%%%%%%%%%%%%%%
\subsubsection{Poynting flux}

\begin{figure}
\begin{center}
\includegraphics[angle=0,trim=1cm 0cm 0.5cm 0.5cm,width=0.95\columnwidth]{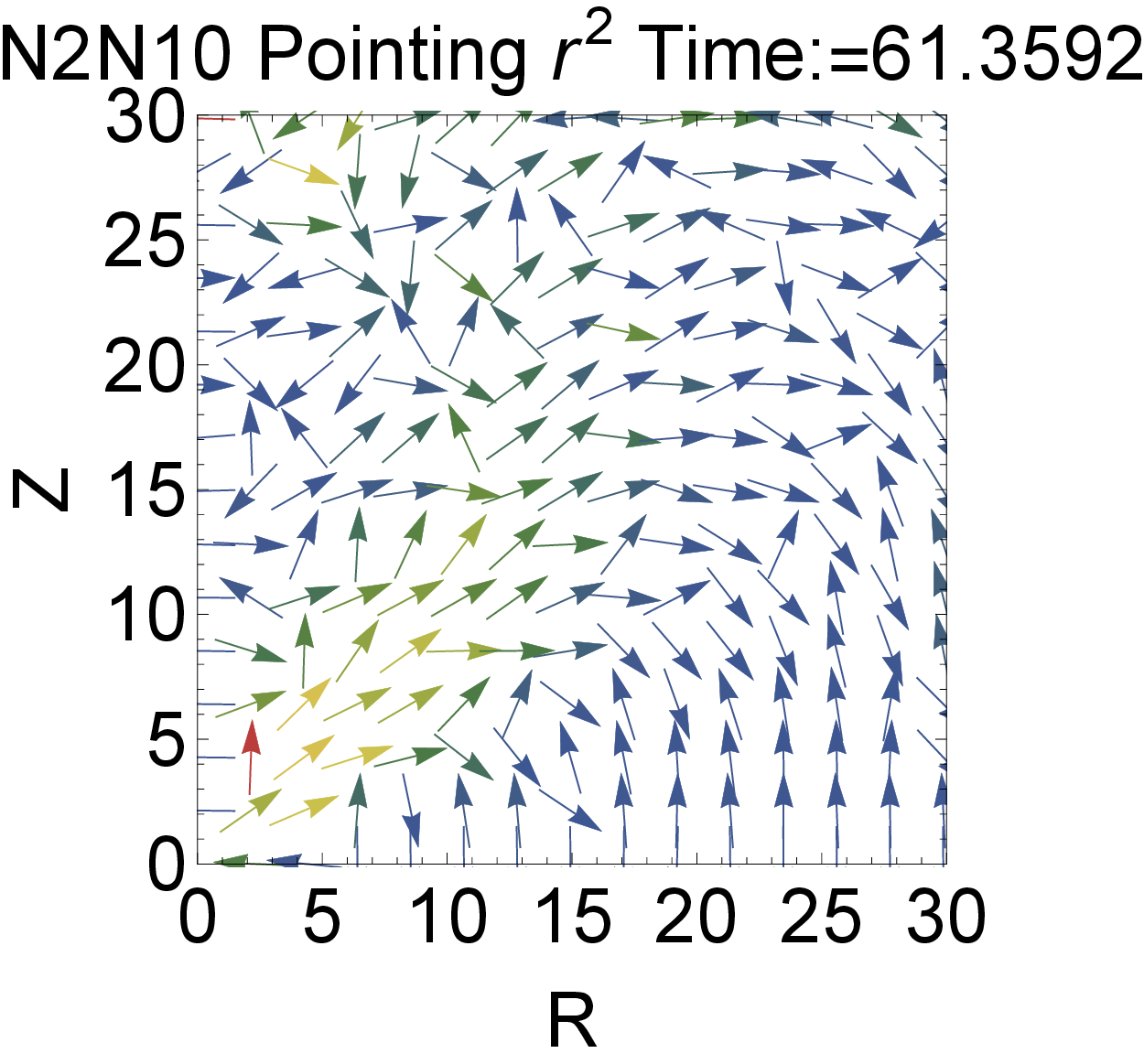}
\caption{For $n = 2, \sigma_{0} = 10$. 
Vector diagram of Poynting flux. The intensities of the vectors increase in the order blue, green, yellow, red.}
\label{pic:n2n10t150point_2.pdf}
\end{center}
\end{figure}

\begin{figure}
\begin{center}
\includegraphics[angle=0,trim=1cm 0cm 0.5cm 0.5cm,width=0.95\columnwidth]{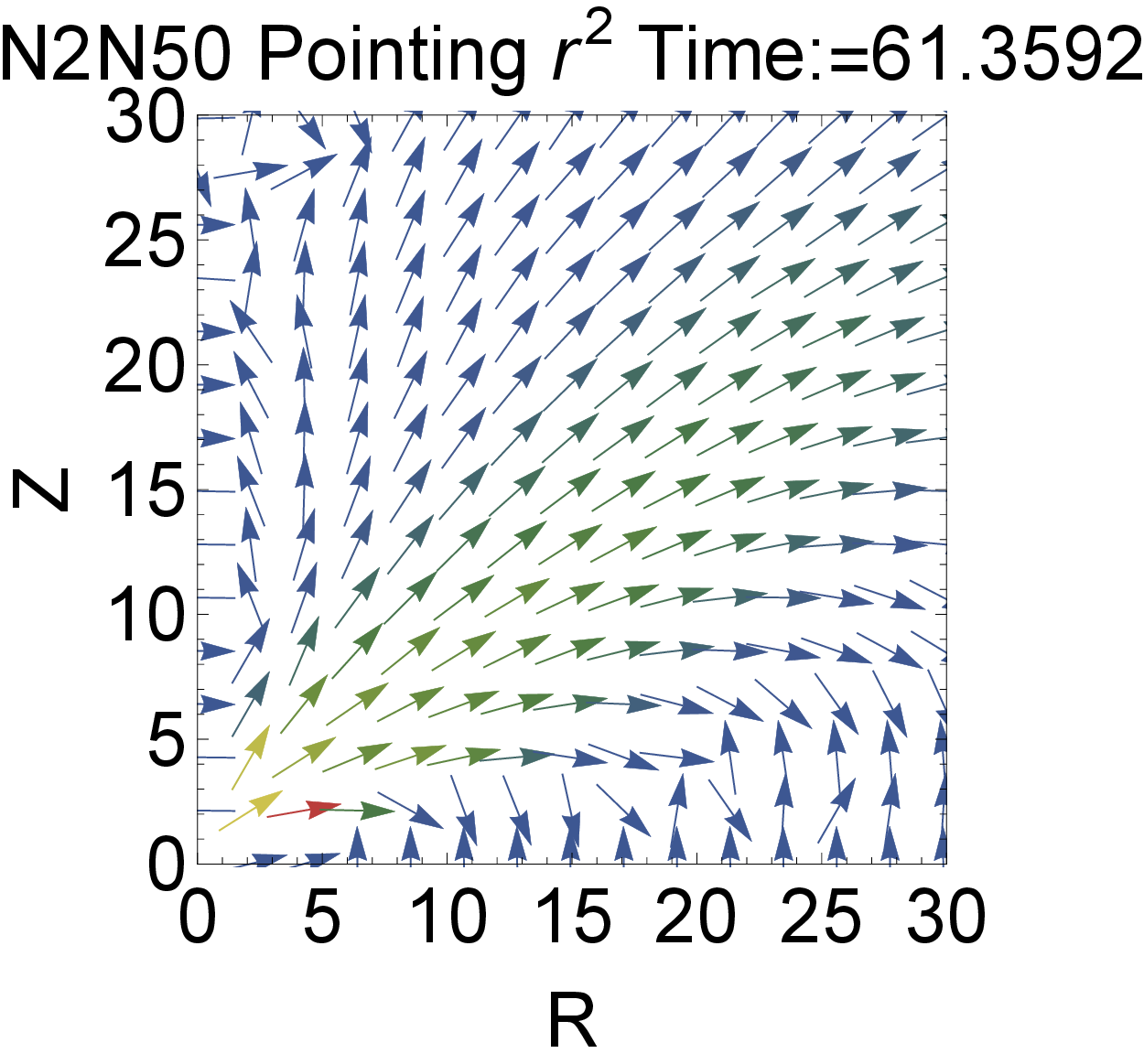}
\caption{Same as Fig. \ref{pic:n2n10t150point_2.pdf} for $n=2,\sigma_{0}=50$}
\label{pic:n2n50t150point_2.pdf}
\end{center}
\end{figure}

Poynting flux is present because there is a nonzero cross product of the electric and magnetic fields; it
Poyntingis emitted from the surface outward.
I calculated the Poynting flux for the obtained steady solution.
Figs. \ref{pic:n2n10t150point_2.pdf} and \ref{pic:n2n50t150point_2.pdf} depict the magnitude and direction distributions of the Poynting vector.
The magnitude of the Poynting vector increases from blue to red.
As shown in Fig. \ref{pic:n2n10st200.pdf}, Poynting flux has a maximum value at the position of the light cylinder, and decreases gradually outside.

The Poynting flux passing through the light cylinder is important.
Thus I calculated the Poynting flux's R component,  $L_{R}(R)$, summing up over Z direction: 

\begin{eqnarray}
L_{R}(R) = 2 \pi R \int^{z \mathrm{max}}_{0} \left( E_{\phi} B_{z} - E_{z} B_{\phi} \right) dz.
\end{eqnarray}

\begin{figure}
\begin{center}
\includegraphics[angle=0,trim=1cm 0cm 0.5cm 0.5cm,width=0.95\columnwidth]{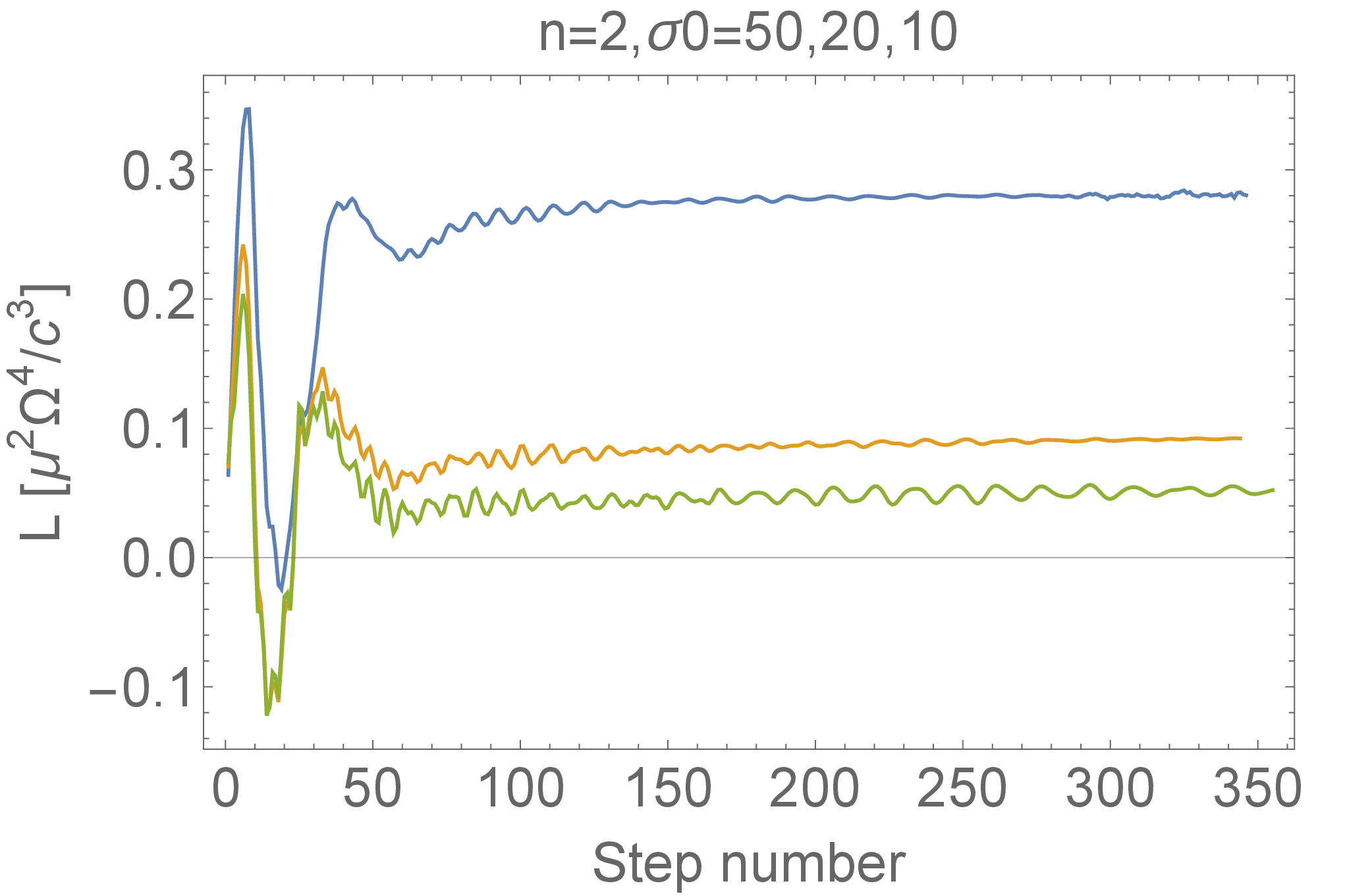}
\caption{The horizontal axis shows the number of steps, and the vertical axis indicates the normalized Poynting flux. Each line can be seen that converges to a constant value at $ \sigma_{0} = 10,20,50 $ for more than time 70 steps, meaning that the magnetosphere is in a stable state over time.}
\label{pic:20141209timesteppointantei.pdf}
\end{center}
\end{figure}

\begin{figure}
\begin{center}
\includegraphics[angle=0,trim=1cm 0cm 0.5cm 0.5cm,width=0.95\columnwidth]{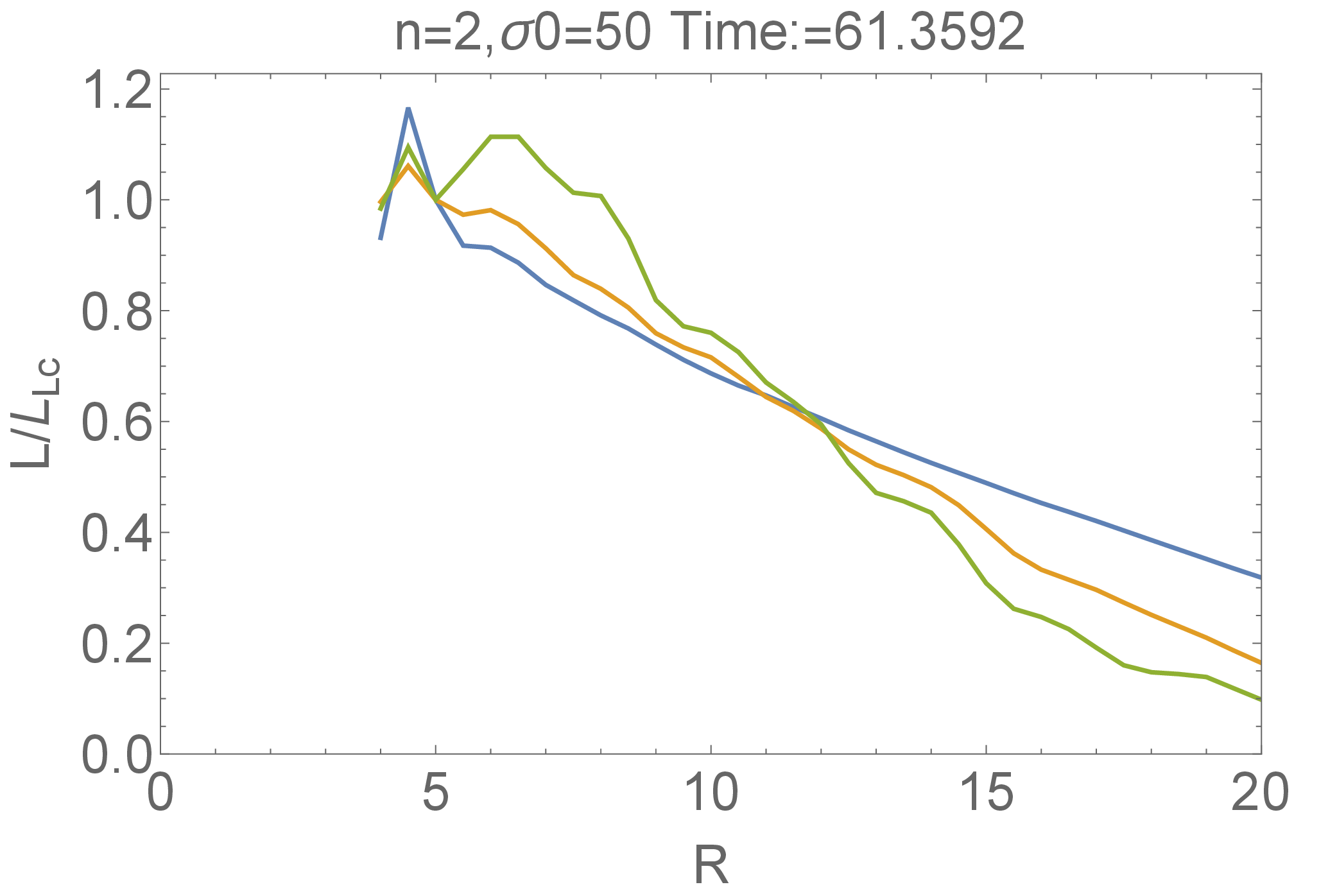}
\caption{The vertical axis Poynting fluxes are normalized by the value of the light cylinder, while the horizontal axis is the distance from the magnetic axis. Poynting flux is maximal at light cylinder $ R=5 $ and is attenuated outside. The green, yellow, and blue lines correspond to $\sigma_ {0} = 10,20,50$. When the electrical conductivity decreases, the Poynting flux damping appear.}
\label{pic:20141125matome150.pdf}
\end{center}
\end{figure}

Figure \ref{pic:20141209timesteppointantei.pdf} shows that the normalized Poynting flux settles  from the inside to the outside over time.
Figure \ref{pic:20141125matome150.pdf} plots the distance dependence of normalized pointing flux on the light cylider.

Poynting flux decreases with radius because the toroidal magnetic field S does not spread to the outside.

In the small conductivity case, Poynting flux decreases rapidly with distance from the surface.
Current circuit structure form within a small region.
Poynting flux increases linearly with respect to electrical conductivity $\sigma_{0}$ in the light cylinder.
However, Poynting flux outside of the light cylinder is not linearly proportional to electrical conductivity $\sigma_{0}$;
this is because the spatial width of the current circuit is changed by the electrical conductivity.
A large spatial scale poloidal current causes a large spatial toroidal magnetic field area.
This is evident in the distribution of the toroidal field S under changing electrical conductivity.

%%%%%%%%%%%%%%%%%%%%%%%%%%%%%%%%%%%%%%%%%%%%%%%%%%%%%%%%%%%%%%%%%%

\subsection{ Electrical conductivity ($\sigma_{0}$) dependence of magnetosphere and Poynting flux }

Under the same field from the equation of current density model,
the current along the magnetic field is increased by high electrical conductivity.
However, it is not a solution that is proportional to the electric conductivity due to its non-linearity.
The most significant change is dependent upon the electrical conductivity $\sigma_{0}$, as it adjusts the spatial scale of the current circuit.
Circuits are smaller when the electrical conductivity $\sigma_{0}$ is small.
Conversely, the high the electric conductivity $\sigma_{0}$, the current circuit becomes large.
As can be seen from the current density distribution, the spatial structure of the current circuit is difficult to grasp clearly.
Therefore, instead of the current density distribution, consider  the toroidal magnetic field distribution.
Fig.4.1 show contours of the toroidal magnetic field S, which is proportional to the electrical conductivity.
Contours have a shape that extends further outward, rather than remaining at the same location.
From the results mentioned above, the current circuit expands by increasing the electrical conductivity.

The contour level is five times larger in Fig.\ref{pic:N2N10_jprS1_00090.pdf}(d) than in Fig.\ref{pic:N2N50_jprS1_00090.pdf}(a).
The maximum toroidal magnetic field S become about five times larger between the $\sigma_{0} =10$ and $\sigma_{0} =50$ cases.
Other $\sigma_{0}$ parameters suggest similar results.

In the near-star region and farther out, the toroidal magnetic-field strength is of the same order and shape.

At around $r > 20$, the toroidal magnetic field S gradually decreases outward.
Magnetic field lines remain open, but current is insufficent to create a toroidal magnetic field.

In Figs. \ref{pic:N2N10_jprS1_00090.pdf},
poloidal current is small in the blue region.
Two characteristic structures appear in Figs.\ref{pic:N2N10_jprS1_00090.pdf}:
the equatorial plane and the closed region.

Radial Poynting flux $L_{R}$ made by $B_{\phi}$ and $E_{z}$ components.
This section discusses the Poynting flux dependence of the electric conductivity. 

\subsubsection{$n=2$ Case}

\begin{figure}
\begin{center}
\includegraphics[angle=0,trim=1cm 0cm 0.5cm 0.5cm,width=0.95\columnwidth]{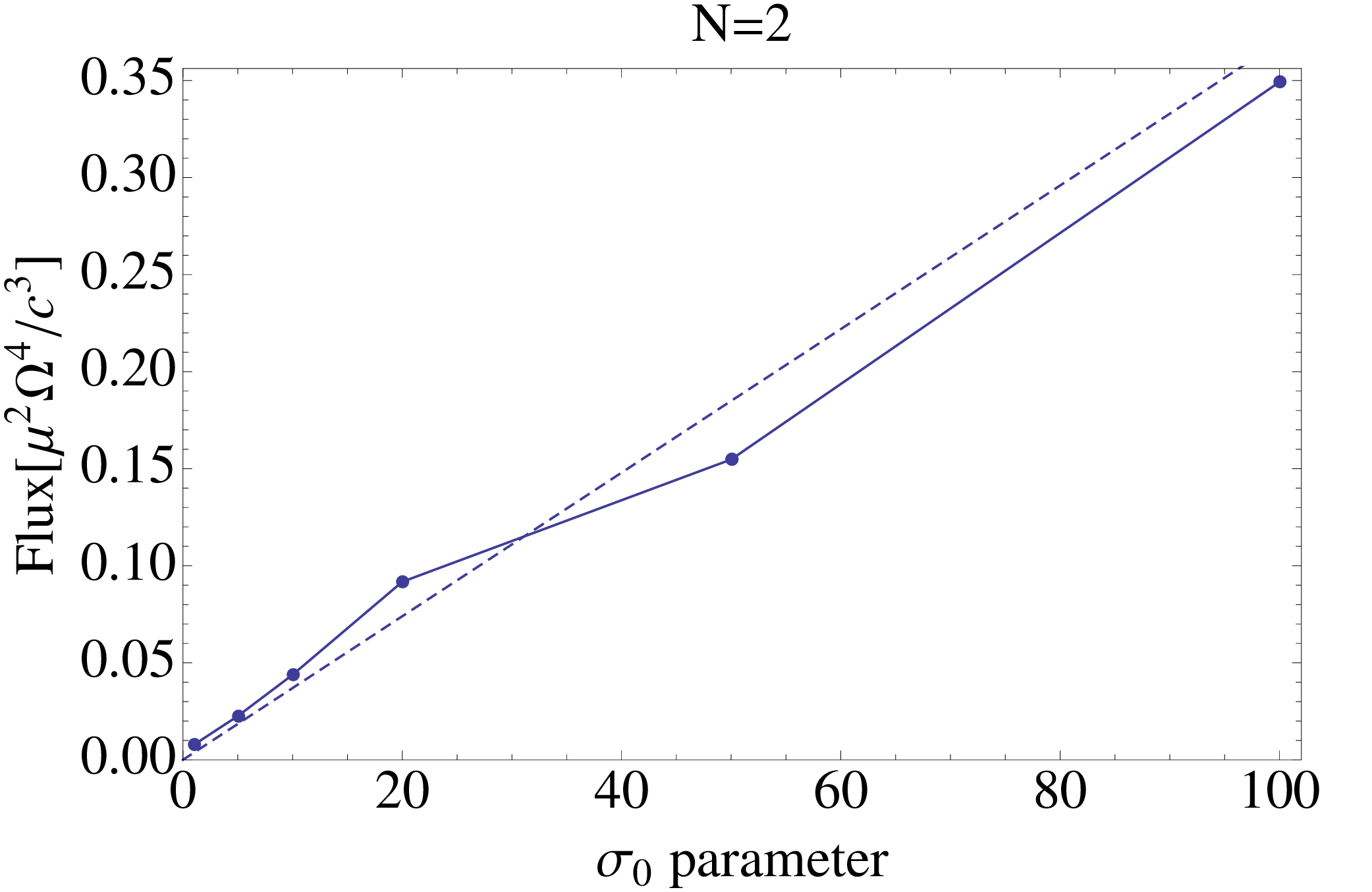}
\caption{Normalized Poynting flux $L /( \mu^{2} \Omega^{4} /c^{3})$ across the light cylinder  of  electric conductivity $\sigma_{0} $ for $n=2$.}
\label{pic:power2pointflux.pdf}
\end{center}
\end{figure}

We set different electric conductivities $\sigma_{0}$ on the star surface,
$\sigma_{0} = \sigma (r=1)$. The radial index is set to $n=2$.
Fig. \ref{pic:power2pointflux.pdf} shows  the regularized Poynting flux to
surface electric conductivity $\sigma_{0}$.
Poynting flux is integrated over the light cylinder radius $R_{\mathrm{LC}}=c/\Omega$, and is proportional to electric conductivity $\sigma_{0}$.

From the fitting result, the linear coefficient is $3.7 \times 10^{-3}  L_{0} / \sigma$, where the normalized Poynting flux is $L_{0} = \mu^{2}\Omega^{4}/c^{3}$.

\begin{table}[htbp]
\caption{Normalized Poynting flux $L /L_{0}$ across the light cylinder  of  electric conductivity $\sigma_{0} $ for $n=2$.}
\begin{center}
\begin{tabular}{ccc}\hline
$\sigma_{0}$ & $\sigma_{0} / \Omega$ & $L / L_{0}$ \\
\hline 
1 & 5 & 0.0078 \\
5 & 25 & 0.0225 \\
10 & 50 & 0.0438 \\
20 & 100 & 0.0917 \\
50 & 250 & 0.1548 \\
100 & 500 & 0.3492 \\
150 & 750 & 0.6032 \\
\hline
\end{tabular}
\end{center}
\label{n2table}
\end{table}%

\subsubsection{$n =1$ Case}

\begin{figure}
\begin{center}
\includegraphics[angle=0,trim=1cm 0cm 0.5cm 0.5cm,width=0.95\columnwidth]{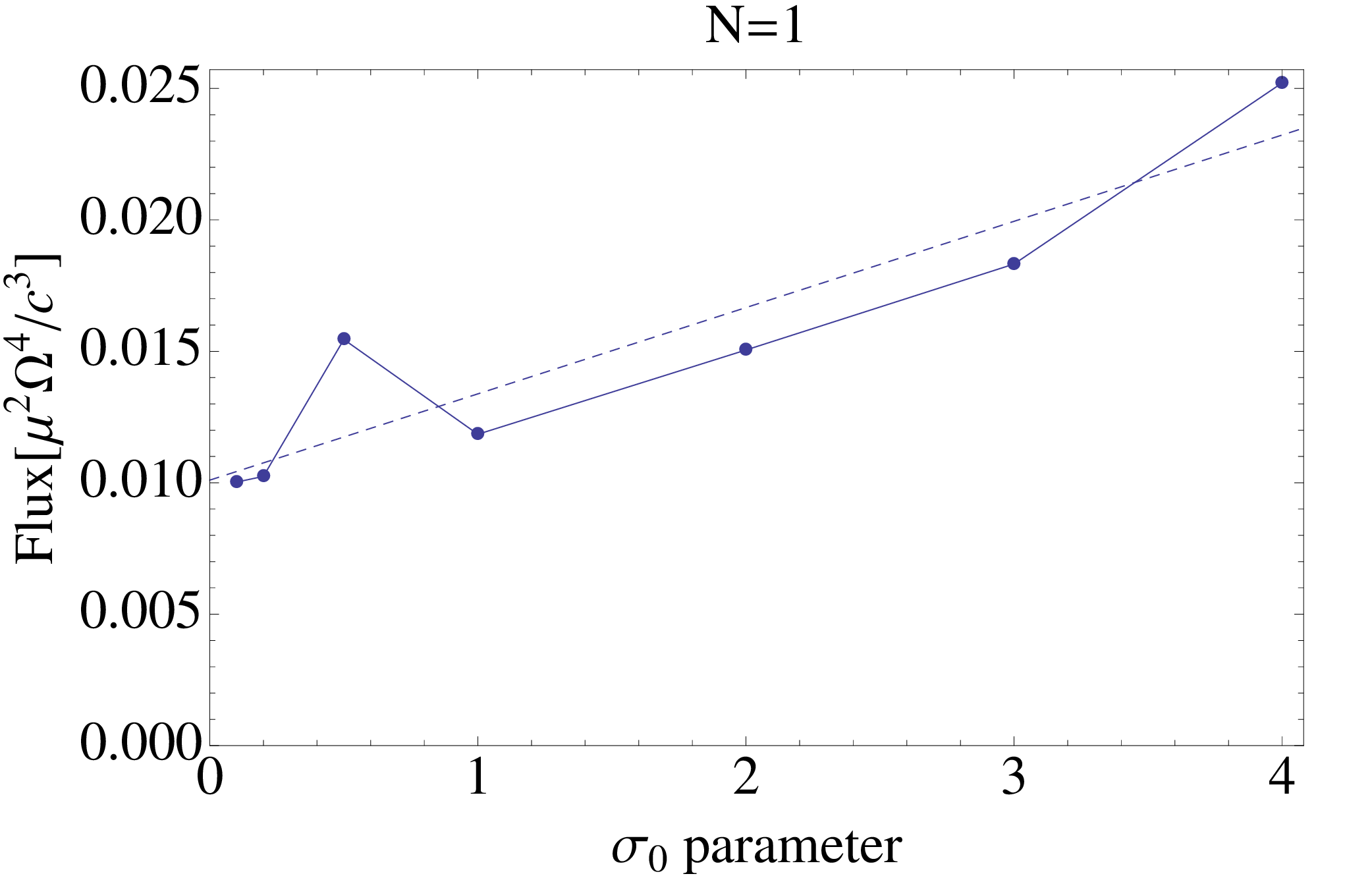}
\caption{Same as Fig. \ref{pic:power2pointflux.pdf}  for $n=1$.}
\label{pic:n1flux.pdf}
\end{center}
\end{figure}

Fig.\ref{pic:power2pointflux.pdf}  and Fig.\ref{pic:n1flux.pdf} show the Poynting flux decrease by radius.
Poynting flux $L$ depend upon electrical conductivity $\sigma_{0}$,
but the Poynting flux in $n=1$ is clearly less than in the $n=2$ case, because the poloidal-current circuit for the $n=2$ case expands more narrowly than for $n=1$.
The Poynting flux for the $n=1$ case is not clearly proportional to the electric conductivity.
 
\begin{table}[htbp]
\caption{Normalized Poynting flux $L /L_{0}$ across the light cylinder  of  electric conductivity $\sigma_{0}$ for $n=1$.}
\begin{center}
\begin{tabular}{ccc}\hline
$\sigma_{0}$ & $\sigma_{0} / \Omega$ & $L / L_{0}$ \\
\hline
0.1 & 0.5 & 0.0100 \\
0.2 & 1.0 &  0.0102\\
0.5 & 2.5 & 0.0155 \\
1 & 5 &  0.0118 \\
2 & 10 & 0.0151 \\
3 & 15 &  0.0183\\
4 & 20 &  0.0252 \\
\hline
\end{tabular}
\end{center}
\label{n1table}
\end{table}%

\section{Discussion}

\subsection{The global magnetospherical structure's electrical-conductivity dependence}
Our simulation introduces the radial dependency the of electric conductivity $\sigma (r)$,  eq. (\ref{sigma0dep}).
$\sigma (r)$ is parameterized by $n$ and $\sigma_{0}$.
The current density is derived from Ohm's law;
however, the current density includes velocity along the magnetic field as a parameter and cannot be determined uniquely.

Poynting flux increases monotonically along with electrical conductivity $\sigma_{0}$ because toroidal magnetic field is increased by a large poloidal current.
In both the n=1 case and n=2 cases, Poynting flux is not linearly dependent on the parameters.
The poloidal current circuit has width and circuit shape.
Poynting flux gradually decreases outward.
In the case of $n = 1$, Poynting flux increases gradually compared to the $n=2$ case.
It is found that in the $n = 1$ case, the current circuit is more widely spread than in the n = 2 case.

To compared to other research results in  the parameter  $ \sigma / \Omega $.
In previous papers, only $\sigma_{0}$ has been considered as a parameter.
Thus, I use the value of $\sigma (r=5) / \Omega$.
The magnitude of the Poynting flux in each of the parameter is different.

From the force-free simulation result by \citet{Spitkovsky2006}, the Poynting flux is

\begin{eqnarray}
\frac{L}{L_{0}} =  \left( 1+ \sin^{2} \alpha \right),
\end{eqnarray}

where $\alpha$ is the angle between the rotation and magnetization axes.
$\alpha=0$ corresponds to an aligned rotator.
Poynting flux is smaller than that in the paper \cite{Li2011}.

\citet[][]{Li2011} shows the Poynting flux in the resistive force-free condition. 
In the $n=2$ case, we have
\begin{eqnarray}
\frac{L}{L_{0}} = -2.3 \times 10^{-3} + 1.9 \times 10^{-2} \left( \frac{\sigma}{\Omega} \right).
\label{li2011eq1}
\end{eqnarray}
In the $n=1$ case, we have
\begin{eqnarray}
\frac{L}{L_{0}} =  1.2 \times 10^{-2} + 8.3 \times 10^{-4} \left( \frac{\sigma}{\Omega} \right).
\end{eqnarray}

\citet[][]{Li2011} corresponds to the case of $n = 0$.
In this case, the electric current is spread more widely than in the $n = 1$ and $n = 2$ cases.
Our simulation outer electric conductivity close to vacuum.
Thus, Poynting flux becomes small.
Poynting flux in the light cylinder is proportional to electrical conductivity. 
It is different from their result.

\subsection{The magnetospheric structure independent of the surface-boundary toroidal magnetic field}

\begin{figure*}
\centerline{
\includegraphics[angle=0,width=0.91\textwidth]{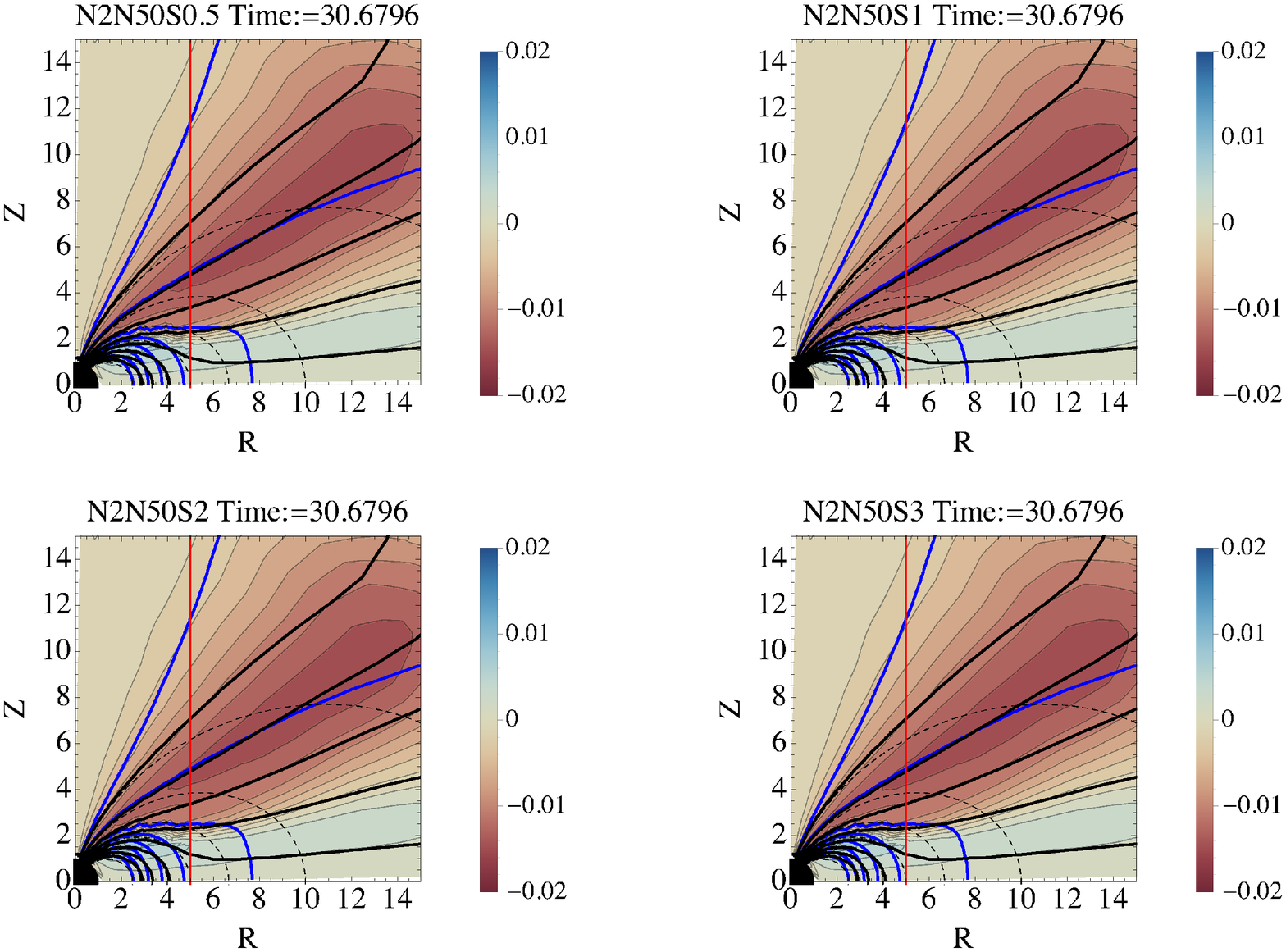}}
\caption{ Poloidal and toroidal magnetic field meridional plane by surface change S. Top to bottom and left to light: $S(r,\theta ) = \alpha_{s} S_{0}(r,\theta )$, parameter  $\alpha_{s} =0.5, 1.0,2.0, 3.0$}
\label{pic:N2N50GS00075.pdf}
\end{figure*}

Current circuits connect to the star surface;
however, the toroidal magnetic field set boundary $B_{\phi} =0 $ eq.(\ref{eq:kyoukai_B}).%(3.15).
To check whether the toroidal magnetic field of the star surface affects the magnetosphere, a calculation was performed under different surface toroidal magnetic field conditions.
The toroidal magnetic field in the surface boundary set  $S(r,\theta ) = \alpha_{s} S_{0}(r,\theta )$, and parameter changed $\alpha_{s} =0.5, 1.0,2.0, 3.0$.
The resulting magnetic field is shown in Fig.\ref{pic:N2N50GS00075.pdf}.
The results show that the Poynting flux does not change for parameter $\alpha_{s}$. 
Poynting flux L is independent of the surface toroidal magnetic field.
The magnetic field formed by the current circuit can be considered independent.

\subsection{Solve both hemispheres in the magnetosphere}

\begin{figure}
\begin{center}
\includegraphics[angle=0,width=\columnwidth]{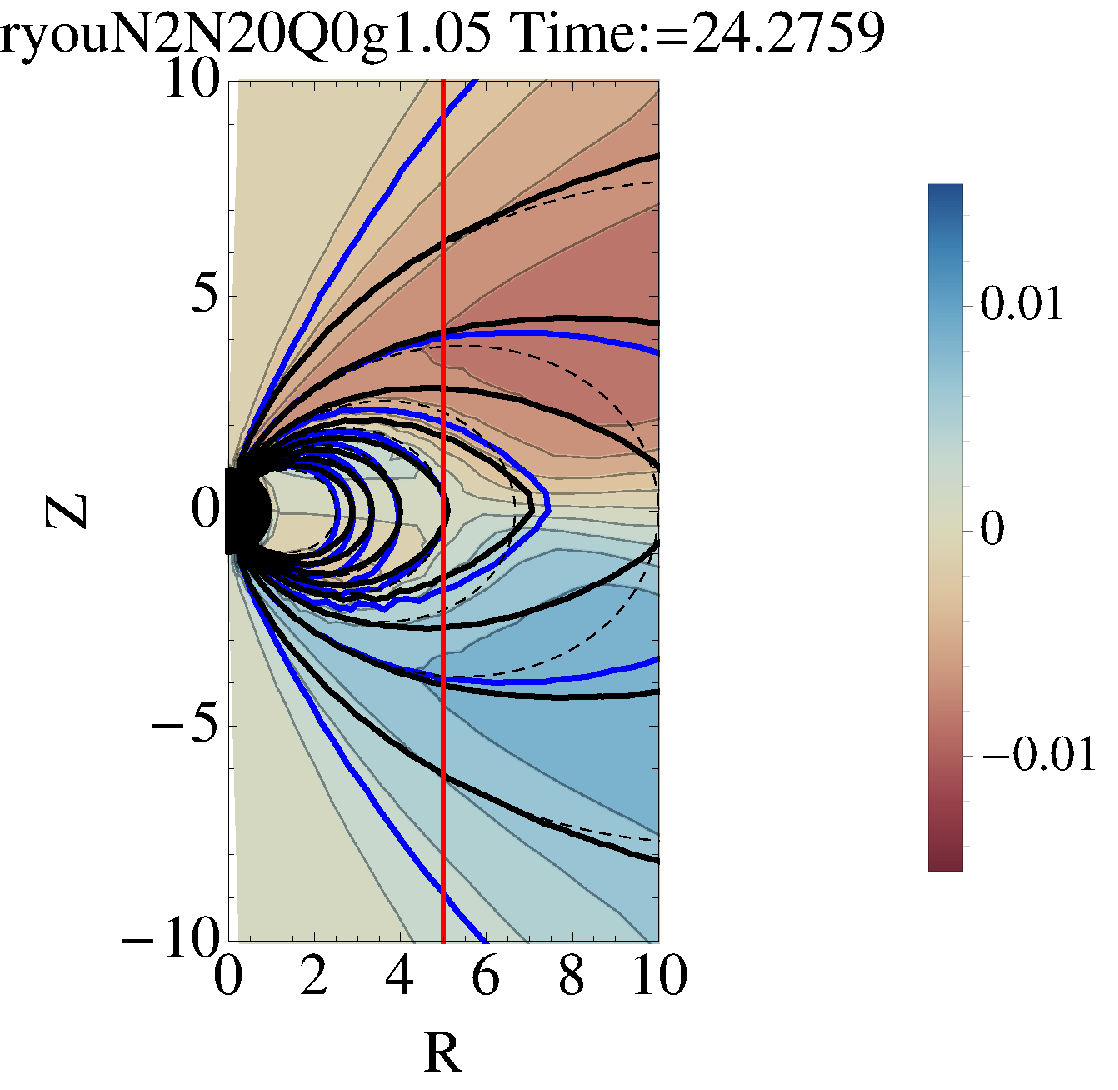}
\caption{Whole spherical simulation result. Same as Fig.\ref{pic:n2n10st200.pdf} for  $n=2$,$\sigma_{0}=20$}
\label{pic:both}
\end{center}
\end{figure}

Solutions that dose not form magnetic field line structures may remain open for a closed configuration.
This result is in contrast to other studies showing that magnetic field lines gradually opened under strong toroidal current flows in the equatorial plane.
In this study, the electrical conductivity in the equatorial plane is set to be the same as in other region.
Fig. \ref{pic:both} shows the results, which predict the magnetosphere grows unstable overtime.
Color represents the toroidal magnetic field. 
Instability was developed in the vicinity of the Y-point and equatorial plane beyond the light cylinder.
When the electrical conductivity is large, instability grows faster.
In the closed magnetic field configuration case, electric conductivity can range up to $\sigma_{0} \sim 60$. 
In the open-boundary-condition case, electrical conductivity can range up to  $\sigma_{0} = 150$.

\subsection{Equatorial plane current sheet}
The equatorial plane boundary condition is set as $E_{\theta}=0$.
The radial magnetic direction is antiparallel to the upper and lower equatorial planes.
In center of the current sheet, the magnetic field is very weak.
There is no boundary condition for $E_{r}$.
Thus $E^{2} > B^{2}$ does not meet the force-free condition.
The force-fee split-monopole and dipole solution has the same character to the current-sheet formation.
The split monopole yields a magnetic flux of $f(Z/R)$. 
From $\mathrm{Ampe^{'}re's}$ law, a poloidal magnetic field open in  infinite distance needs a toroidal current sheet in the equatorial plane. 
The equatorial toroidal current sheet is described using the Dirac delta function.
In the dipole magnetic field case, electric current concentrates in the last open field line. In this paper, the equatorial plane boundary condition is idealized to solve problems.

%%%%%%%%%%%%%%%%%%%%%%%%%%%%%%%%%%%%%%%%%%%%%%%%%%%%%%%%
\section{Conclusion}

I use a current model derived from Ohm's law to understand resistive force-free magnetospheres.
I introduce an electrical conductivity dependent upon distance from the star.
A steady state is obtained by  combining Maxwell equations and the boundary condition. 
These resistive force-free solutions show that the current has width and circuit shape.
A toroidal magnetic field is formed outside of the light cylinder.
The Poynting flux from magnetosphere has a maximum in the light cylinder and decreases on the outside.  
In the high-conductivity case, the current circuit spreads more widely.  
The shape of the current circuit varies with the spatial dependence of the electrical conductivity.
The Poynting flux in the light cylinder is in proportion to the electrical conductivity.
The surface toroidal magnetic field does not affect Poynting flux in the light cylinder.

\section*{Acknowledgment}
The author is thankful to Yasufumi Kojima for many useful comments and their help with the code.

\bibliographystyle{apj}
\bibliography{astro}

\end{document}